\documentclass[twoside,12pt]{article}
\usepackage{epsfig}
\usepackage{amsmath}
\usepackage{amssymb}

\usepackage{graphicx,color}
\def\vec#1{\mbox{\boldmath $#1$}}

\begin{document}
%\draft
%\preprint{}

\title{ Alpha-Particle Condensation  
in Nuclear Systems}
\author{Y. Funaki$^1$, T. Yamada$^2$, H. Horiuchi$^{3,4}$, 
G. R\"opke$^5$, P. Schuck$^{6,7}$ and A. Tohsaki$^3$ \\
$^1$Nishina Center for Accelerator-Based Science, \\
The Institute of Physical and Chemical Research (RIKEN), Wako 351-0198, Japan \\
$^2$Laboratory of Physics, Kanto Gakuin University, Yokohama 236-8501, Japan \\
$^3$Research Center for Nuclear Physics (RCNP), Osaka University, \\
Ibaraki, Osaka 567-0047, Japan \\
$^4$ International Institute for Advanced Studies, Kizugawa 619-0225, Japan \\
$^5$Institut f\"ur Physik, University of Rostock, Universit\"atsplatz 1, \\
18051 Rostock, Germany \\
$^6$Institut de Physique Nucl\'eaire, CNRS, UMR8608, Orsay, F-91406, France\\
$^7$Universit\'e Paris-Sud, Orsay, F-91505, France \\
}

%\date{\today}

\maketitle

\begin{abstract}
  The onset of quartetting, i.e. $\alpha$-particle condensation, in symmetric 
nuclear matter is studied with the help of an in-medium modified four nucleon 
equation. It is found that at very low density quartetting wins over pairing, 
because of the strong binding of the $\alpha$-particles. The critical 
temperature can reach values up to around 6 MeV. Also the disappearance of 
$\alpha$-particles with increasing density, i.e. the Mott transition, is 
investigated. In finite nuclei the Hoyle state, that is the ${0_2}^+$ of 
$^{12}$C, is identified as an ``$\alpha$-particle condensate'' state. It is 
conjectured that such states also exist in heavier $n\alpha$-nuclei, like 
$^{16}$O, $^{20}$Ne, etc. For instance the 6-th $0^+$ state of $^{16}$O at 15.1 MeV is identified from a 
theoretical analysis as being a strong candidate for an $\alpha$ 
condensate state.
  Exploratory calculations are performed for the
  density dependence of the $\alpha$ condensate fraction at zero
  temperature to address the suppression of the four-particle
  condensate below nuclear-matter density. Possible quartet condensation in 
other systems is discussed briefly.\\
\noindent 
Keywords: nuclear matter, $\alpha$-matter, superfluidity, Bose-Einstein
condensation, strongly coupled systems
\end{abstract}

%\pacs{PACS numbers: 05.30.Fk}

\section{Introduction}

One of the most amazing phenomena in quantum many-particle systems is
the formation of quantum condensates.  At present, the formation of
condensates is of particular interest in strongly coupled fermion
systems in which the crossover from Bardeen-Cooper-Schrieffer (BCS)
pairing to Bose-Einstein condensation (BEC) may be investigated. Among
very different quantum systems such as the electron-hole-exciton system in
excited semiconductors, atoms in traps at extremely low temperatures, etc.,
nuclear matter is especially well suited for the study of correlation 
effects in a quantum liquid.

Neutron matter, nuclear matter, but also finite nuclei are superfluid.
However, at low density, nuclear matter will not cluster into pairs, i.e.
deuterons but rather into $\alpha$ -particles which are much more stable. Also
heavier clusters, starting with Carbon, may be of importance but are 
presently not considered for condensation phenomena. Therefore, one may ask 
the question whether there exists quartetting, i.e. $\alpha$-particle 
condensation, in nuclei, analogous to nuclear pairing. The only nucleus 
which in its ground state
has a pronounced $\alpha$ -cluster structure is $^8$Be. In section 4 we will
show a figure of $^8$Be in the laboratory frame and in the intrinsic deformed
frame. We will see that $^8$Be is formed out of two $\alpha$-particles
roughly 4 fm apart, only mildly interpenetrating one another. Actually $^8$Be is 
slightly unstable and the two $\alpha$'s only hold together via the Coulomb 
barrier.
Because of the large distance of the two $\alpha$-particles, the 0$^+$ ground
state of $^8$Be has, in the laboratory frame, a spherical density
distribution whose average is very low: about 1/3 of ordinary saturation
density $\rho_0$. $^8$Be is, therefore, a very large object with an rms 
radius of
about 3.7 fm to be compared with the nuclear systematics of $R = r_0A^{1/3}$ = 
2.44 fm. Definitely $^8$Be is a rather unusual and, in its kind, unique 
nucleus. One may ask the question what happens when one brings a third 
$\alpha$-particle alongside of $^8$Be. We know the answer: the 3-$\alpha$ 
system collapses to the ground state of $^{12}$C which is much denser than 
$^8$Be and can not accommodate with its small radius of 2.4 fm three more or 
less free $\alpha$-particles barely touching one another. One nevertheless may 
ask the question whether the dilute three $\alpha$ configuration $^8$Be-$\alpha$, or rather $\alpha$-$\alpha$-$\alpha$, may not form an isomeric or 
excited 
state of $^{12}$C. That such a 
state indeed exists will be one of the main subjects of our considerations.
Once one accepts the idea of the existence of an $\alpha$-gas state in 
$^{12}$C, there is no reason why equivalent states at low density should not 
also exist in heavier $n\alpha$-nuclei, like $^{16}$O, $^{20}$Ne, etc.
The possible existence of a loosely bound 4$\alpha$ state in $^{16}$O will be 
another topic of our presentation.
 In a 
mean field picture, i.e. all $\alpha$'s being ideal bosons ( in this context 
remember that the first excited state of an $\alpha$-particle is at 
$\sim$ 20 MeV, by factors higher than in all other nuclei), all $\alpha$'s 
will 
occupy the lowest $0S$-state, i.e. 
they will condense. This forms, of course, not a macroscopic condensate but 
it can be understood in the same sense as we know that nuclei are superfluid 
because of the presence of a finite number of Cooper pairs. On the other hand, 
for example during the cooling process of compact stars \cite{ShapT}, where 
one predicts 
the presence of $\alpha$-particles \cite{ST}, a real macroscopic phase of 
condensed $\alpha$'s may be formed. In the present contribution we will mainly 
concentrate on nuclear systems but we also can think about the possibility of 
quartetting in other Fermi-systems. One should, however, keep in mind that a 
pre-requisite for its existence is, as in nuclear physics, that there are four 
different types of fermions. For example to form quartets with cold atoms one 
could try to trap fermions in four different magnetic substates, a task which 
eventually seems possible \cite{Salo}. Also theoretical works in this direction have appeared in the mean while \cite{Lecheminant,miyake}. The fact that the $\alpha$-particle condensates in nuclei do not form the ground state, may give raise to questions. In this respect, one should note that Bose condensates of cold atoms in traps also are not in their ground state which is a solid. It is a question of time scales. $\alpha$-particle condensate states in nuclei usually live four orders of magnitude longer than typical nuclear times.

In the next section we will investigate how the binding energy of various 
nuclear clusters change with density. In section 3 we study the critical 
temperature of $\alpha$-particle condensation in infinite matter via an in-medium four-nucleon equation (Thouless criterion) and in section 4 we treat 
$\alpha$-particle condensation in finite nuclei. In section 5 we give our results for $^{12}$C and in section 6 the ones for $^{16}$O. In section 7, we briefly discuss the question of the occupation 
numbers of the $\alpha$-particles in the various nuclear states with emphasis 
on the condensate states. In section 8 we present a simplified calculation 
of the condensate fraction in $\alpha$-matter. Finally in section 9 we conclude with outlook and further discussions.

\section{Nuclear clusters in the medium}

With increasing density of nuclear matter, medium modifications of
single-particle states as well as of few-nucleon states become of
importance. The self-energy of an $A$-particle cluster can in principle
be deduced from contributions describing the single-particle
self-energies as well as medium modifications of the interaction and
the vertices. A guiding principle in incorporating medium effects is
the construction of {\it consistent} (``conserving'') approximations, which
treat medium corrections in the self-energy and in the interaction
vertex at the same level of accuracy. This can be achieved in a
systematic way using the Green functions formalism \cite{KKER}.  At the 
mean-field
level, we have only the Hartree-Fock self-energy $\Gamma^{\rm HF} =
\sum_2 V(12,12)_{\rm ex} f(2)$ together with the Pauli blocking
factors, which modify the interaction from $V(12,1'2')$ to
$V(12,1'2')[1 - f(1) - f(2)]$, with 
$f(1)=[1+\exp(E^{\rm HF}(1)-\mu)/T]^{-1}$.  
In the case of the two-nucleon system
($A=2$), the resulting effective wave equation which includes those 
corrections reads
\begin{equation}
\label{two_part_bind}
\left[E^{\rm HF}(1)+E^{\rm HF}(2)-E_{2,n,P}\right] \psi_{2,n,P}(12) +
\sum_{1'2'}[1-f(1)-f(2)]\,\,V(12,1'2') 
 \psi_{2,n,P}(1'2')=0.
\end{equation}
This {\it effective wave equation} describes bound states as well as
scattering states. The onset of pair condensation is achieved when the 
binding energy $E_{d,P=0}$ coincides with $2 \mu$.

Similar equations have
been derived from the Green function approach for the case $A = 3$ and $A = 4$,
describing triton/helion ($^3$He) nuclei
as well as $\alpha$-particles in nuclear matter. The effective wave
equation contains in mean field approximation the Hartree-Fock
self-energy shift of the single-particle energies as well as the Pauli
blocking of the interaction. We give the effective wave equation for
$A=4$, 
\begin{eqnarray}
\label{four_part_bind}
&& \left[E^{\rm HF}(1)+E^{\rm HF}(2)+ E^{\rm HF}(3)+E^{\rm HF}(4)
  -E_{4,n,P}\right] 
\psi_{4,n,P}(1234) \nonumber\\ && + 
\sum_{i<j}\sum_{1'2'3'4'}[1-f(i)-f(j)]V(ij,i'j')\prod_{k\neq i,j}\delta_{k,k'} 
 \psi_{4,n,P}(1'2'3'4')=0.
\label{EWE}
\end{eqnarray}
A similar equation is obtained for $A=3$. 

\begin{figure}[h]
\hspace{0.4cm}
\begin{minipage}[h]{8.5cm}
\psfig{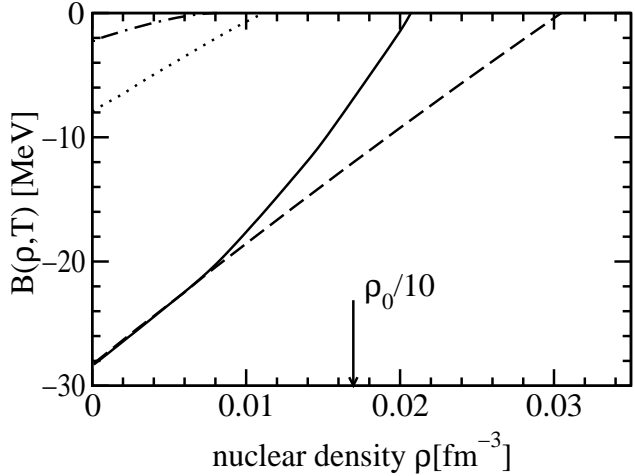}
\end{minipage}
\hfill
\hspace{-0.4cm}
\begin{minipage}[h]{8.5cm}
\caption{Shift of binding energy of the light clusters ($d$ - dash dotted, 
$t/h$ - dotted, and $\alpha$ - dashed: perturbation theory, full line : 
non-perturbative Faddeev-Yakubovski equation) in symmetric nuclear
matter as a function of density for given temperature $T = 10$ MeV 
\cite{Beyer}.}
\label{shifts}
\end{minipage}
\end{figure}

The effective wave equation has been solved using separable potentials
for $A=2$ by integration. For $A=3,4$ we can use a {\it Faddeev
approach} \cite{Beyer}.  The shifts 
of binding energy can also be calculated approximately via perturbation 
theory.  In Fig.~\ref{shifts} we show the shift of the binding 
energy of the light clusters ($d, t/h$ and $\alpha$) in 
symmetric nuclear matter as a function of density for 
temperature $T$ = 10 MeV. 

It is found that the cluster binding energy decreases 
with increasing density.  Finally, at the {\it Mott density} 
$\rho_{A,n,P}^{\rm Mott}(T)$ the bound state is dissolved.  
The clusters are not present at higher densities,  merging into the 
nucleonic medium.  For a given
cluster type characterized by $A,n$, we can also introduce 
the Mott momentum $P^{\rm Mott}_{A,n}(\rho,T)$ in terms of
the ambient temperature $T$ and nucleon density $\rho$,
such that the bound states exist only for 
$P \ge P^{\rm Mott}_{A,n}(\rho,T)$.
 We do not present an example here, but it is intuitively clear that a 
cluster with high c.o.m. momentum with respect to the medium is less 
affected by the Pauli principle than a cluster at rest.

\section{Four-particle condensates and quartetting in nuclear matter}

In general, it is necessary to take into account of {\it all bosonic clusters} to
gain a complete picture of the onset of superfluidity.
As is well known, the deuteron is weakly bound as compared to
other nuclei.  Higher $A$-clusters can arise that are more stable.  In
this section, we will consider the formation of $\alpha$-particles,
which are of special importance because of their large binding energy
per nucleon ($\sim 7$ MeV).  We will not include tritons or helions, which
are fermions and not so tightly bound.  Moreover, we will not consider
nuclei in the iron region, which have even larger binding energy per
nucleon than the $\alpha$-particle and thus constitute, in principle, 
the dominant component at
low temperatures and densities. However, the latter are complex structures of
many particles and are strongly affected by the medium as the density
increases for given temperature, so that they are assumed not to be of relevance in the
density region considered here.

The in-medium wave equation for the four-nucleon problem has been 
solved using the Faddeev-Yakubovski technique, with the inclusion of
Pauli blocking.  The binding energy of an $\alpha$-like cluster 
with zero c.o.m.\ momentum vanishes at around $\rho_0/10$, where
$\rho_0 \simeq 0.16$ nucleons/fm$^3$ denotes the saturation density 
of isospin-symmetric nuclear matter, see Fig.~\ref{shifts}.  Thus, 
the four-body bound 
states make no significant contribution to the composition of 
the system above this density.  Given the medium-modified 
bound-state energy $E_{4,P}$, the bound-state contribution to 
the EOS is
\begin{equation}
\rho_4(\beta,\mu) = \sum_P\left[e^{\beta(E_{4,P} 
- 2 \mu_p-2 \mu_n)} -1\right]^{-1}\,.
\end{equation}
We will not include the contribution of the excited states or
that of scattering states.  Because of the large specific 
binding energy of the $\alpha$ particle, low-density nuclear 
matter is predominantly composed of $\alpha$ particles.
This observation underlies the concept of $\alpha$ matter
and its relevance to diverse nuclear phenomena.

\begin{figure}[t]
\hspace{0.7cm}
\begin{minipage}[bth]{7.5cm}
\psfig{figure=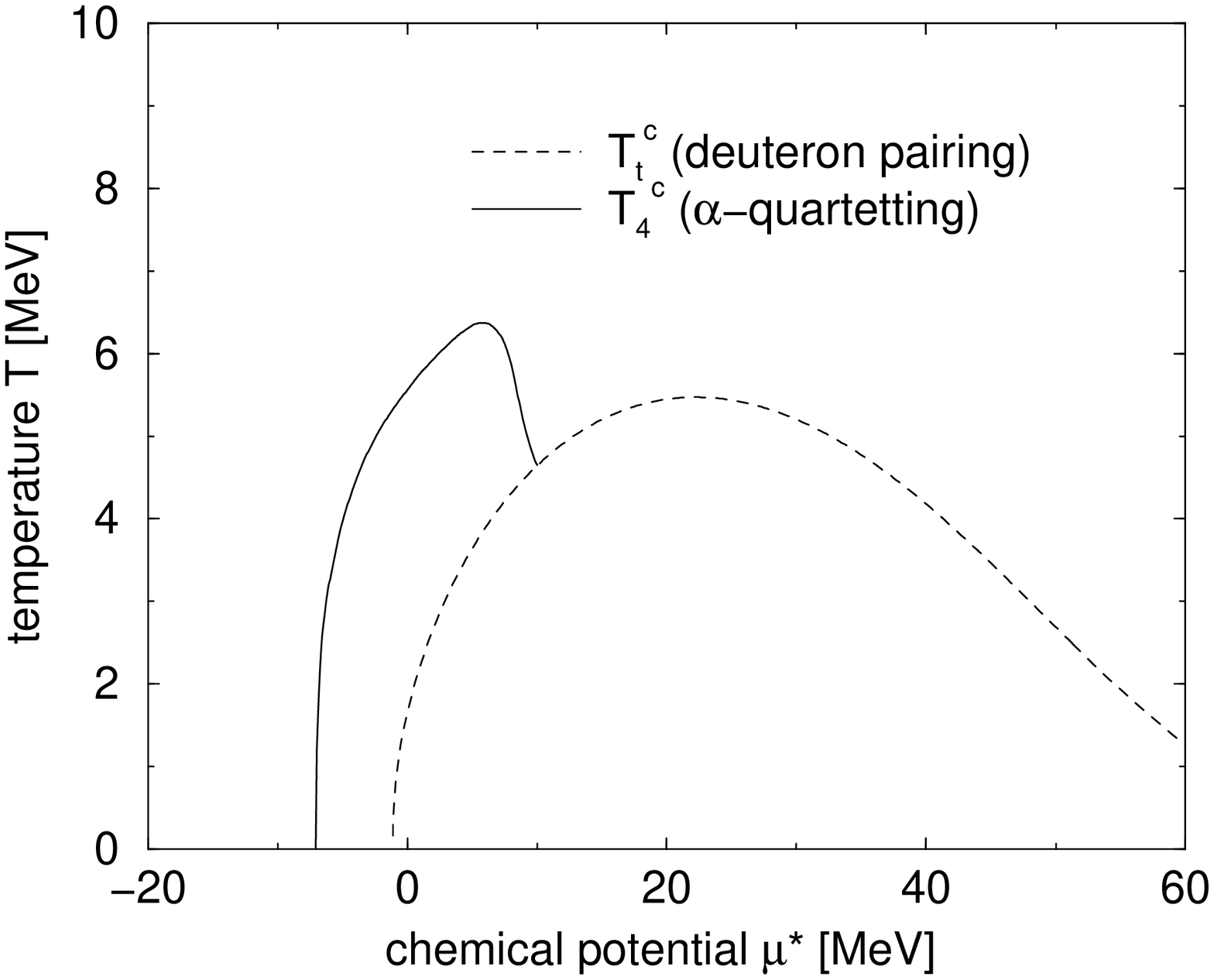,width=7.cm}
\caption{Transition temperature to quartetting/pairing as a function of
  chemical potential in symmetric nuclear matter.}
\label{fig:trans_mu}
\end{minipage}
\hfill
\hspace{-0.7cm}
\begin{minipage}[bth]{9.cm}
\psfig{figure=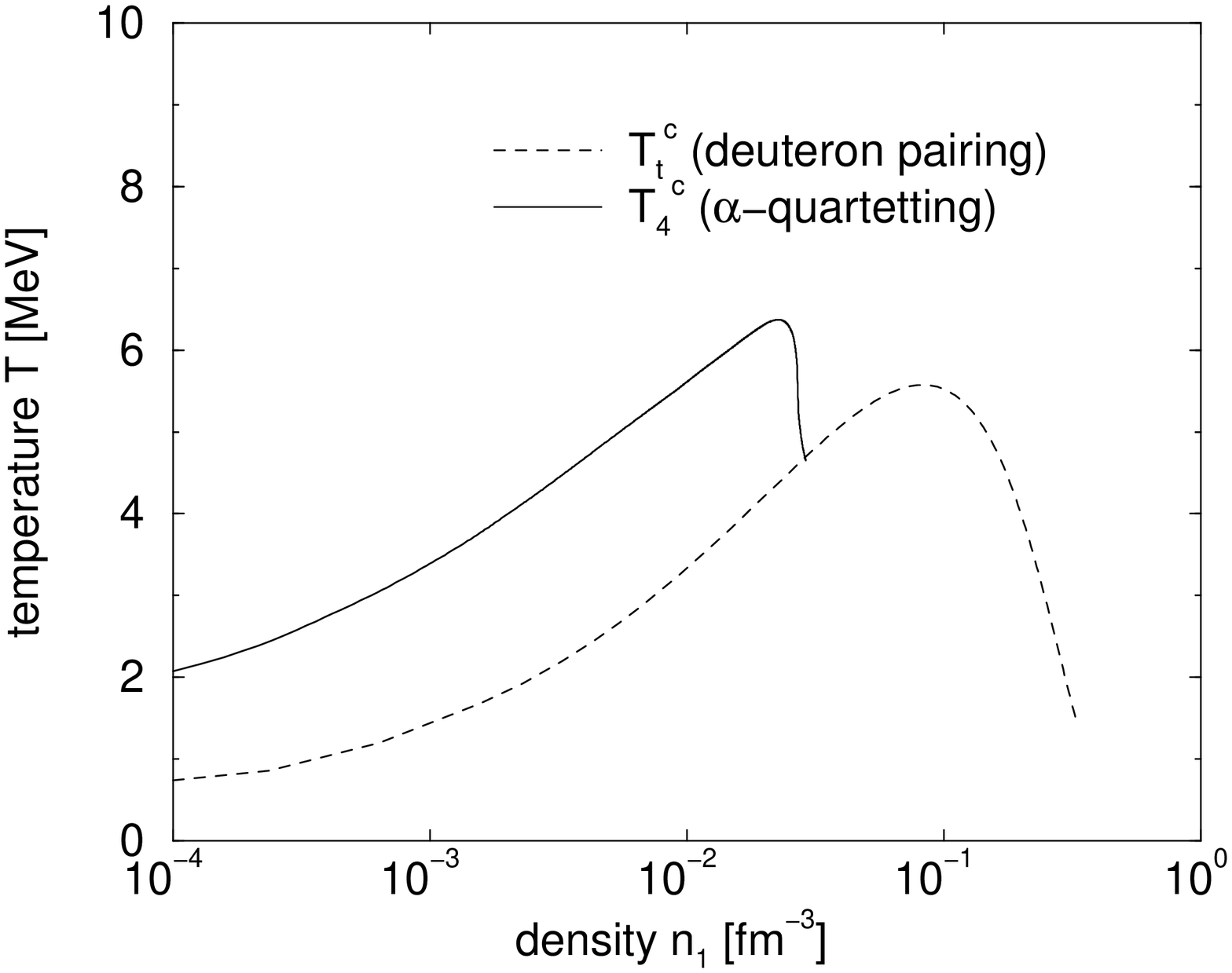,width=7.3cm}
\caption{Transition temperature to quartetting/pairing as a function of
nucleon density in symmetric nuclear matter.}
\label{fig:trans_dens}
\end{minipage}
\end{figure}

As exemplified by Eq.~(\ref{EWE}), the 
effect of the medium on the properties of an $\alpha$ particle in
mean-field approximation (i.e., for an uncorrelated medium) is 
produced by the Hartree-Fock self-energy shift and Pauli blocking.
The shift of the $\alpha$-like bound state has been calculated using
perturbation theory \cite{RMS} as well as by solution of the
Faddeev-Yakubovski equation \cite{Beyer}. 
%It is found that this bound state merges with the continuum of scattering states at a Mott density $\rho_\alpha^{\rm Mott} \approx \rho_0/10$, see Fig.~\ref{shifts}. 
It is found that the bound states of clusters $d$, $t$,
and $h$ with $A<4$ are already dissolved at a Mott density
$\rho_\alpha^{\rm Mott} \approx \rho_0/10$, see Fig.~\ref{shifts}. Since Bose condensation only is of relevance for $d$ and $\alpha$, and the fraction of $d$, $t$ and $h$ becomes low compared with that of $\alpha$ with increasing density, we can neglect the contribution of them to an equation of state. 
Consequently, if we further neglect the 
contribution of the four-particle scattering phase shifts 
in the different channels, we can now construct an equation of state 
$\rho(T, \mu) =\rho^{\rm free}(T, \mu) + 
\rho^{{\rm bound}, d}(T, \mu) 
+\rho^{{\rm bound}, \alpha}(T, \mu)$ such that
$\alpha$-particles determine the behavior of symmetric nuclear matter
at densities below $\rho_\alpha^{\rm Mott}$ and temperatures below 
the binding energy per nucleon of the $\alpha$-particle. The 
formation of deuteron clusters alone
gives an incorrect description because 
the deuteron binding energy is small, and the abundance of
$d$-clusters is small compared with that of $\alpha$-clusters. 
In the low density region of the phase diagram, $\alpha$-matter emerges as
an adequate model for describing the nuclear-matter equation
of state.

With increasing density, the medium modifications -- especially Pauli
blocking -- will lead to a deviation of the critical temperature
$T_c(\rho)$ from that of an ideal Bose gas of $\alpha$-particles
(the analogous situation holds for deuteron clusters, i.e., in 
the isospin-singlet channel).

 Symmetric nuclear matter is characterized by the
equality of
the proton and neutron chemical potentials,
i.e., $\mu_p=\mu_n=\mu$. 
Then an extended Thouless condition based on the relation for the four-body 
T-matrix (in principle equivalent to Eq.~(\ref{EWE}) at eigenvalue 4$\mu$)
\begin{eqnarray}
{\rm T}_4(1234,1''2''3''4'', 4 \mu)& =& \sum_{1'2'3'4'} \Biggl\{
  \frac{V(12,1'2')[1-f(1)-f(2)] }{ 4 
    \mu - E_1-E_2-E_3-E_4 }\delta(3,3')\delta(4,4')\nonumber\\ 
&& \qquad \qquad + {\rm cycl.} \Biggr\}
{\rm T}_4(1'2'3'4',1''2''3''4'', 4 \mu)
\end{eqnarray}
serves to determine the onset of Bose condensation of $\alpha$-like
clusters, noting that the existence of a solution of this relation signals
a divergence of the four-particle correlation function. 
An approximate solution has been obtained by a variational
approach, in which the wave function is taken as Gaussian 
incorporating the correct solution for the two-particle
problem \cite{RSSN}.

The results are presented in Figs.~\ref{fig:trans_mu} 
and \ref{fig:trans_dens}. An important consequence of those is that at the 
lowest 
temperatures, 
Bose-Einstein condensation occurs for $\alpha$ particles rather
than for deuterons.  As the density increases within the low-temperature
regime, the chemical potential $\mu$ first reaches $-7$ MeV, where 
the $\alpha$'s Bose-condense.  By contrast, Bose condensation
of deuterons would not occur until $\mu$ rises to $-1.1$ MeV.

The {\it ``quartetting''} transition temperature sharply 
drops as the rising density approaches the critical Mott 
value at which the four-body bound states disappear.  At that point,
pair formation in the isospin-singlet deuteron-like channel 
comes into play, and a deuteron condensate will exist below the 
critical temperature for BCS pairing up to densities above
the nuclear-matter saturation density $\rho_0$, as described in 
the previous Section.  
The critical density at which the $\alpha$ 
condensate disappears is estimated to be $\rho_0/3$. Therefore, $\alpha$-particle condensation primarily only exists in the Bose-Einstein-Condensed 
(BEC) phase and there does not seem to exist a phase where the quartets 
acquire a large extension as Cooper pairs do in the weak coupling regime.  
However, 
the variational approach of Ref.~\cite{RSSN} on which 
this conclusion is based represents only a first attempt at the
description of the transition from quartetting to pairing.  
The detailed nature of this fascinating transition remains 
to be clarified.

Many different questions arise in relation to the possible
physical occurrence and experimental manifestations of quartetting: 
Can we observe the hypothetical ``$\alpha$ condensate'' in nature?  
What about thermodynamic stability?  What happens with quartetting 
in asymmetric nuclear matter?  Are more complex quantum 
condensates possible?  What is their relevance for finite 
nuclei?  As discussed below, the special type of microscopic 
quantum correlation associated with quartetting may be important 
in nuclei, its role in these finite inhomogeneous systems being 
similar to that of pairing.

\section{Description of Alpha-Particle Condensate States in Self-Conjugate 4n Nuclei}

\begin{figure}[h]
\hspace{0.7cm}
\begin{minipage}[h]{8.cm}
\psfig{figure=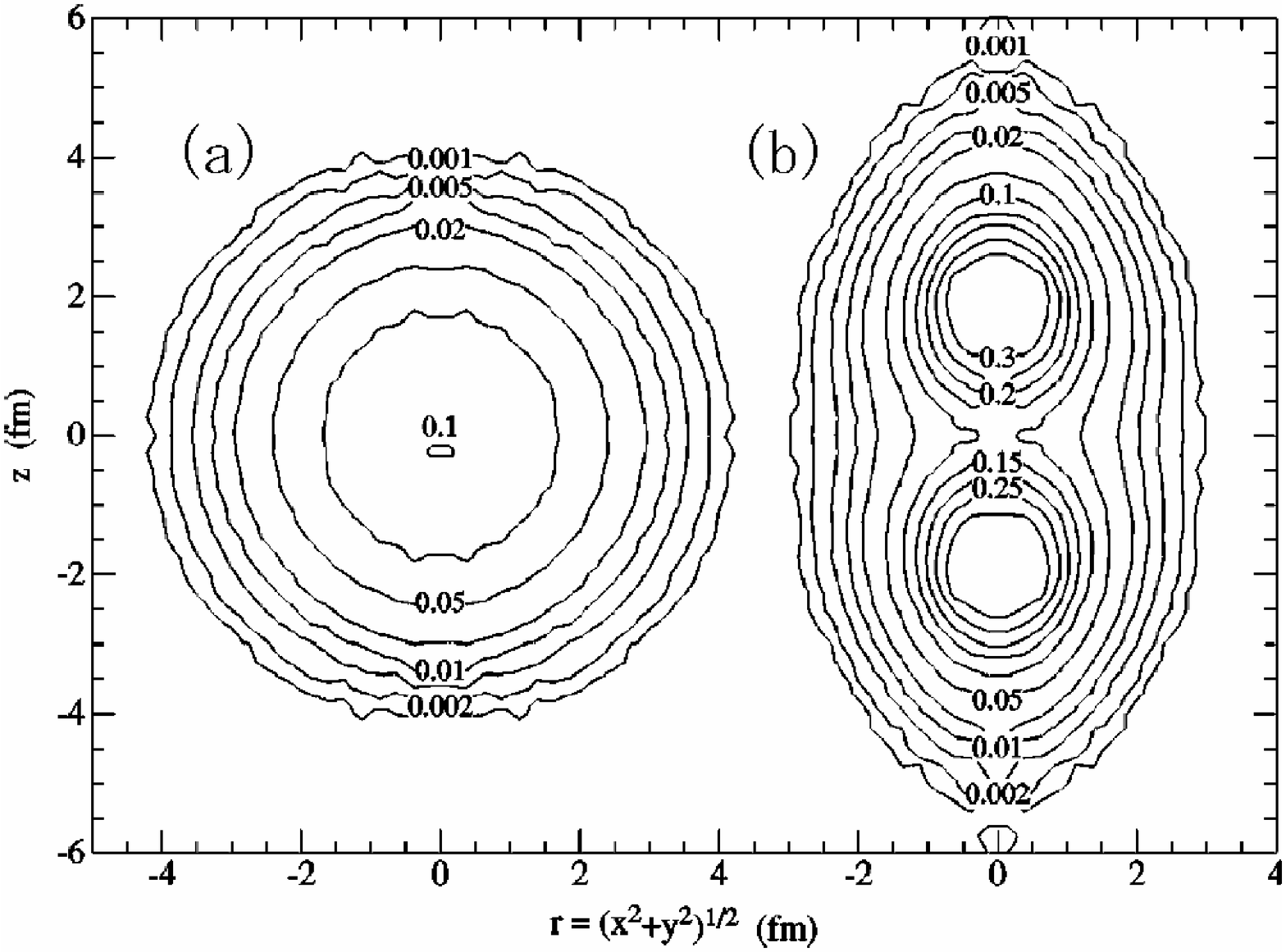,width=7.5cm}
\end{minipage}
\hfill
\hspace{-0.5cm}
\begin{minipage}[h]{9.5cm}
\caption{Contours of constant density (taken from Ref.~\cite{qmc}),
 plotted in cylindrical coordinates, for $^8$Be$(0^+)$. The left
 side (a) is in the ``laboratory'' frame while the right side (b) is in
 the intrinsic frame.}
\label{profiles}
\end{minipage}
\end{figure}

Let us discuss the possibility of quartetting in nuclei. The 
only nucleus having a pronounced $\alpha$-cluster structure 
in its ground state is $^8$Be.  In Fig.~\ref{profiles}(a), we 
show the result of an exact calculation of the density 
distribution of $^8$Be in the laboratory frame.  In 
Fig.~\ref{profiles}(b) we show, for comparison, the result
of the same calculation in the intrinsic, deformed frame.  We see a 
pronounced two $\alpha$-cluster structure where the two $\alpha$'s are 
$\sim$ 4 fm apart, giving rise 
to a very low average density $\rho \sim \rho_0/3$
as seen in Fig.~\ref{profiles}(a).  As already discussed in the introduction, 
$^8$Be is a rather unusual and unique
nucleus. One may be intrigued by the question, already raised earlier,  
whether 
loosely bound $\alpha$-particle configurations may not also exist in heavier $n\alpha$-nuclei, at 
least in excited states, naturally close to the $n\alpha$ disintegration 
threshold. Since $\alpha$-particles are rather inert bosons ( first excited 
state at $\sim$ 20 MeV), these $\alpha$-particles then would all condense in 
the lowest $S$-wavefunction, very much in the same way as do bosonic atoms in 
magneto-optical traps \cite{String}. This
question and exploring related issues of quartetting
in finite nuclei will consume most of the rest of the present 
study.

In fact, we will be able to offer strong arguments that the 
$0_2^+$ state of $^{12}$C at 7.654 MeV is a state of 
$\alpha$-condensate nature.  Later, also indications for the existence of an 
analogous state in $^{16}$O will be discussed.

First, it should be understood that the $0_2^+$ state in 
$^{12}$C is in fact hadronically unstable (as $^8$Be), being 
situated about 300 keV above the three $\alpha$-break up threshold. 
This state is stabilized only by the Coulomb barrier. 
It has a width of $8.7$ eV and a corresponding 
lifetime of $7.6\times 10^{-17}$ s.  As well known, this state is
of paramount astrophysical (and biological!) importance due to 
its role in the creation of $^{12}$C in stellar nucleosynthesis. 
Its existence was predicted in 1953 by the astrophysicist Fred 
Hoyle \cite{hoyle}. His  prediction was confirmed experimentally 
a few years later by Willy Fowler and coworkers at Caltech \cite{fowler}. 
It is also well known that this {\it Hoyle state}, as it is now called, 
is a notoriously difficult state for any nuclear theory to explain.
For example, the most modern no-core shell-model 
calculations predict the $0_2^+$ state in $^{12}$C to lie
at around 17 MeV above the ground state -- more than 
twice the actual value \cite{nocore}.  This fact alone tells 
us that the Hoyle state must have a very unusual structure.  
It is easy to understand that, should it indeed have the proposed 
loosely bound three $\alpha$-particle structure, a shell-model 
type of calculation would have great difficulties in reproducing 
its properties. 

An important development bearing on this issue took place some 
thirty years ago.  Two Japanese physicists, M. Kamimura \cite{kamimura} 
and K. Uegaki \cite{uegaki}, along with their collaborators, 
almost simultaneously reproduced the Hoyle state from a microscopic 
theory.  They employed a twelve-nucleon wave function together 
with a Hamiltonian containing an effective nucleon-nucleon 
interaction.  At that time, their work did not attract 
 the attention it deserved; the true importance of their 
achievement has been appreciated only recently.  The two groups 
started from practically the same ansatz for the $^{12}$C wave 
function, which has the following three $\alpha$-cluster structure: 
%#JWC - going to a display equation -
\begin{equation}
\langle \vec{r}_1...\vec{r}_{12}|^{12}{\rm C} \rangle = 
{\cal{A}}\left[\chi({{\vec{s}}},\vec{t})\phi_1\phi_2\phi_3\right]\,. 
\end{equation}
In this expression, the operator ${\cal A}$ imposes antisymmetry
in the nucleonic degrees of freedom and $\phi_i$, with $i=1,2,3$ for 
the three $\alpha$'s, is an intrinsic $\alpha$-particle wave function of 
prescribed Gaussian form, 
\begin{equation}
\phi(\vec{r}_1,\vec{r}_2,\vec{r}_3, \vec{r}_4) = 
\exp\left\{-\left[(\vec{r}_1-\vec{r}_2)^2
+(\vec{r}_1-\vec{r}_3)^2 +\cdots \right]/{8b^2}\right\}\,, 
\label{alphawf}
\end{equation}
where the size parameter $b$ is adjusted to fit the rms of 
the free $\alpha$-particle, and $\chi({{\vec{s}}},\vec{t})$ 
is a yet-to-be determined three-body wave function for the c.o.m.\
motion of the three $\alpha$'s, their corresponding Jacobi coordinates 
being denoted by ${\vec{s}}$ and $\vec{t}$.  The unknown 
function $\chi$ was determined via calculations
based on the Generator Coordinate Method \cite{uegaki} (GCM) 
and the Resonating Group Method \cite{kamimura} (RGM) calculations 
using the Volkov I and Volkov II 
nucleon-nucleon forces, which fit $\alpha$-$\alpha$ phase 
shifts. The precise solution of this complicated three body problem,
carried out three decades ago, was truly a pioneering achievement,
with results fulfilling expectations. The position of the Hoyle state, 
as well as other properties including the inelastic form factor and 
transition probability, successfully reproduced the experimental 
data.  Other states of $^{12}$C below and around the energy of 
the Hoyle state were also successfully described.  Moreover, it 
was already recognized that the three $\alpha$'s in the Hoyle state 
form sort of a gas-like state.  In fact, this feature had previously 
been noted by H.~Horiuchi \cite{hori} prior to the appearance
of Refs.~\cite{kamimura,uegaki}, based on results from the
orthogonality condition model (OCM) \cite{saitoh}. 
All three Japanese research groups concluded from their studies 
that the linear-chain state of three $\alpha$-particles, 
postulated by Morinaga many years earlier \cite{mori} as an interpretation 
of the Hoyle state, had to be rejected.

Although the evidence for interpreting the Hoyle state 
in terms of an $\alpha$ gas was stressed in the cited papers
from the late 1970's, two important aspects of the situation 
were missed at that time.  First, because the three $\alpha$'s 
move in identical $S$-wave orbits, one is dealing with an
$\alpha$-condensate state, albeit not in the macroscopic sense, and that 
this may be a quite general phenomenon, also in heavier self-conjugate nuclei. 
The second  important point is that the complicated three-body wave 
function $\chi({\vec{s}},{\vec{t}})$ for the c.o.m. motion of the three 
$\alpha$'s 
can be replaced by a structurally and conceptually very 
simple microscopic three-$\alpha$ wave function of the 
condensate type, which has practically 100 percent overlap 
with the previously constructed ones \cite{thsr} \cite{cbec} (see also 
Ref.~\cite{Hackenbroich}). We now 
describe this condensate wave function.

\begin{figure}[htbp]
\begin{center}
\includegraphics[scale=1.]{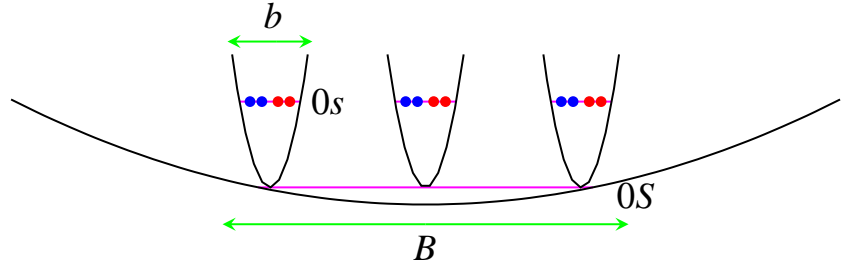}
\caption{Pictorial representation of the THSR wave function for $n=3$ ($^{12}$C). The three $\alpha$-particles are trapped in the $0S$-state of a wide harmonic oscillator $(B)$ and the four nucleons of each $\alpha$ are confined in the $0s$-state of a narrow one $(b)$. All nucleons are antisymmetrised.}
\label{fig:osc}
\end{center}
\end{figure}

We start by examining the BCS wave function of ordinary fermion
pairing, obtained by projecting the familiar BCS 
ground-state ansatz onto an $N$-particle subspace of Fock space.  
In the position representation, this wave function is
\begin{equation}
\langle \vec{r}_1\cdots \vec{r}_N|{\rm BCS}\rangle = {\cal {A} }
\left[\phi(\vec{r}_1,\vec{r}_2)\phi(\vec{r}_3,\vec{r}_4)\cdots \phi(\vec{r}_{N-1}
\vec{r}_N)\right]\,, \label{eq:1}
\end{equation}
where $\phi(\vec{r}_1,\vec{r}_2)$ is the Cooper-pair wave function 
(including spin and isospin), which is to be determined variationally  
through the familiar BCS equations.  The condensate character of the BCS 
ansatz is borne out by the fact that within the antisymmetrizer $\cal A$,
one has a product of $N/2$ times the same pair wave function $\phi$, 
with one such function for each distinct pair in the reference
partition of $\{1,2,\ldots,N-1,N\}$.  Formally, it is now a simple 
matter to generalize (\ref{eq:1}) to quartet or $\alpha$-particle 
condensation. 
We write
\begin{equation}
\langle \vec{r}_1,\ldots,\vec{r}_N|\Phi_{n\alpha}\rangle = {\cal {A} }\left[ 
\phi_{\alpha}(\vec{r}_1, \vec{r}_2, \vec{r}_3, \vec{r}_4)\phi_{\alpha}
(\vec{r}_5,\ldots , \vec{r}_8)
\cdots \phi_{\alpha}(\vec{r}_{N-3},\ldots, \vec {r}_N)\right]\,, \label{eq:2}
\end{equation}
where $\phi_{\alpha}$ is the wave 
function common to all condensed $\alpha$-particles. Of course, 
finding the variational solution for this function
is, in general, extraordinarily more complicated than finding
the Cooper pair-wave function $\phi$ of Eq.~(\ref{eq:1}). 
Even so, in the present case that the $\alpha$-particle is the
four-body cluster involved, and for applications to 
relatively light nuclei, the complexity of the problem can be reduced 
dramatically.  This possibility stems from the fact, already known 
to the authors of Refs.~\cite{kamimura,uegaki}, that, due to the BEC-character
of $\alpha$-particle condensation (see above), 
an excellent variational ansatz for the intrinsic wave function
of the $\alpha$-particle is provided [as in Eq.~(\ref{alphawf})], 
by a Gaussian form with only the size parameter $b$ to be determined. 
In addition -- and here resides the essential point of our wave function -- even the c.o.m. motion of 
the system of $\alpha$-particles can be described very well 
by a Gaussian wave function with, this time, a size parameter 
$B \gg b$ to account for the motion over the nuclear space.  
We therefore write
\begin{equation}
\phi_\alpha(\vec{r}_1,\vec{r}_2,\vec{r}_3,\vec{r}_4) = 
e^{{\displaystyle{-2}}{\scriptstyle\vec{R}^2}{\displaystyle{/B^2}}}
\phi(\vec{r}_1-\vec{r}_2,\vec{r}_1-\vec{r}_3,\cdots)\,,  \label{eq:3}
\end{equation}
where $\vec{R}= (\vec{r}_1+\vec{r}_2+\vec{r}_3+\vec{r}_4)/4$ is the c.o.m.\ 
coordinate of one $\alpha$-particle and $\phi(\vec{r}_1-\vec{r}_2,...)$ is 
the same intrinsic $\alpha$-particle wave function of Gaussian form as 
already used in Refs.~\cite{kamimura,uegaki} and given explicitly
in Eq.~(\ref{alphawf}).
Naturally, in Eq.~(\ref{eq:2}) the center of mass $\vec{X}_{\rm cm}$ of 
the three $\alpha$'s, i.e., of the whole nucleus, should be eliminated; 
this is easily achieved by replacing $\vec{R}$ by $
\vec{R}-\vec{X}_{\rm cm}$ in 
each of the $\alpha$ wave functions in Eq.~(\ref{eq:2}). 
The 
$\alpha$-particle condensate wave function specified by Eqs.~(\ref{eq:2}) 
and (\ref{eq:3}), proposed in Ref.~\cite{thsr} and henceforth 
called the THSR wave function, now depends on only two parameters, 
$B$ and $b$. The wave function (\ref{eq:2}) with (\ref{eq:3}) is pictorially represented in Fig.~\ref{fig:osc}. The expectation value of an assumed microscopic Hamiltonian
$H$,
\begin{equation}
{\cal {H}}(B,b)=\frac{\langle\Phi_{n\alpha}(B,b)|H|\Phi_{n\alpha}(B,b)\rangle}
{\langle \Phi_{n\alpha}|\Phi_{n\alpha}\rangle}\,, \label{eq:4}
\end{equation}
can be evaluated, and the corresponding two-dimensional energy surface 
can be quantized using the two parameters $B$ and $b$ 
as Hill-Wheeler coordinates. \index{Hill-Wheeler method}

\begin{figure}[htbp]
\begin{center}
\includegraphics[scale=0.7,angle=270]{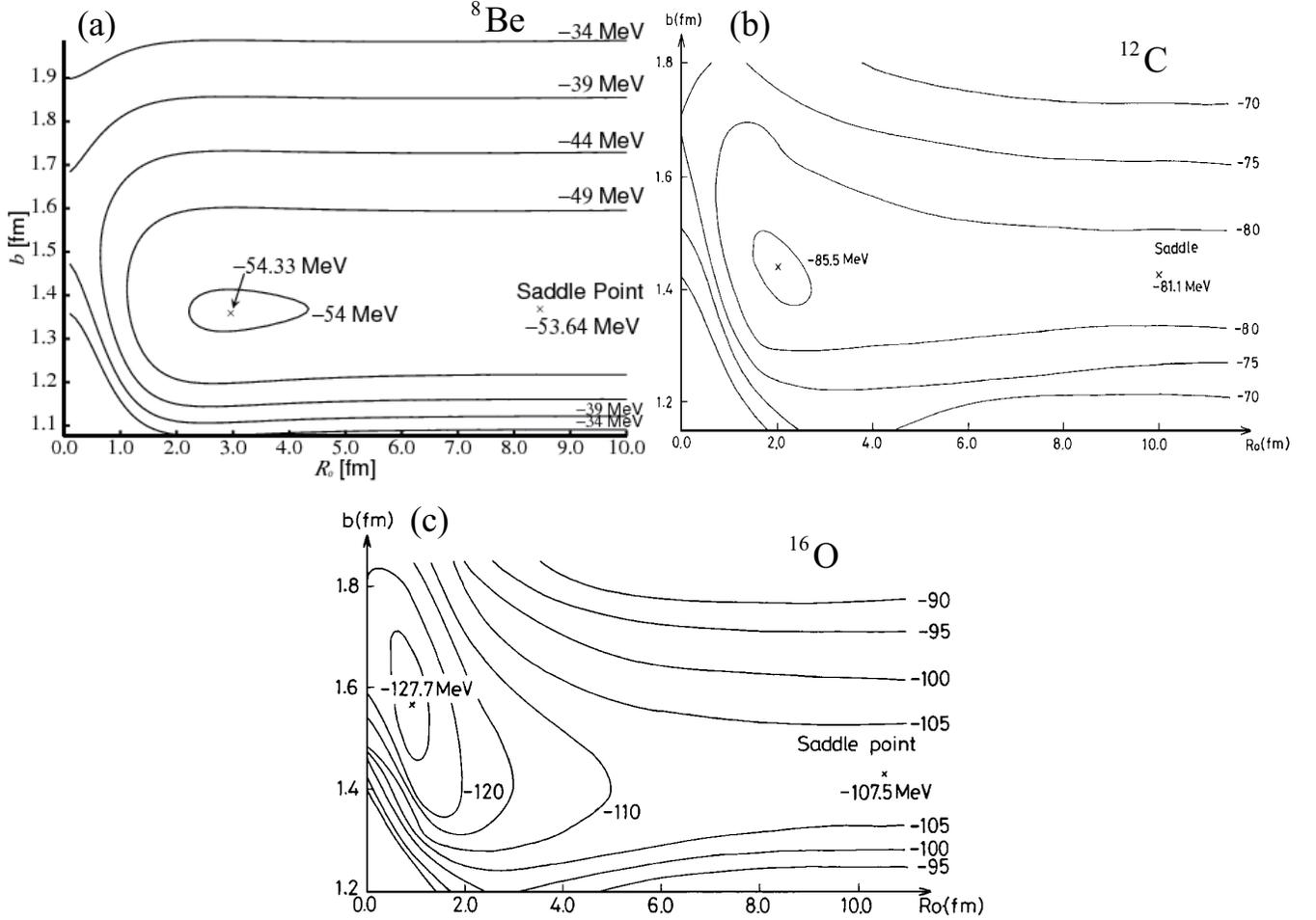}
\caption{Contour map of the energy surface ${\cal {H}}(B,b)$ for (a) $^8$Be, (b) $^{12}$C and (c) $^{16}$O. The variables $B$ and $R_0$ are connected by the relation, $B^2=b^2+2R_0^2$. Numbers attached to the contour lines are the binding energies in units of MeV.}
\label{fig:E_surface}
\end{center}
\end{figure}

Before presenting the results, let us discuss the THSR wave 
function in somewhat more detail.  This innocuous-looking variational
ansatz, namely Eq.~(\ref{eq:2}) together with Eq.~(\ref{eq:3}), 
is actually more subtle than it might at first appear. One should 
realize that two limits are incorporated exactly.  One is obtained by 
choosing $B=b$, for which Eq.~(\ref{eq:2}) reduces to a standard 
Slater determinant with harmonic-oscillator single-nucleon 
wave functions, leaving the oscillator length $b$ as the 
single adjustable parameter.  This holds because the 
right-hand-side of expression (\ref{eq:3}), with $B=b$, becomes 
a product of four identical Gaussians, and the antisymmetrization 
creates all the necessary $P$, $D$, etc.\ harmonic oscillator 
wave functions automatically \cite{thsr}.  On the other hand,
when $B \gg b$, the density of $\alpha$-particles is very low, 
and in the limit $B \rightarrow \infty$, the average distance between 
$\alpha$'s is so large that the antisymmetrisation between 
them can be neglected, i.e., the operator $\cal {A}$ in front of 
Eq.~(\ref{eq:2}) becomes irrelevant and can be removed.
In this limiting case, our wave function then describes an ideal gas of 
independent, condensed $\alpha$-particles -- it is a pure product 
state of $\alpha$'s! An elucidating study on this aspect is given in 
Ref.~\cite{yamada1}.\\
Evidently, however, in realistic cases the antisymmetrizer 
$\cal {A}$ cannot be neglected, and evaluation of the expectation 
value (\ref{eq:4}) becomes a nontrivial analytical task. 
The Hamiltonian in Eq.~(\ref{eq:4}) was taken to be the one
used in Ref.~\cite{tohsaki_F1}, which features an effective 
nucleon-nucleon force of the Gogny type, with parameters fitted
to $\alpha$-$\alpha$ scattering phase shifts as available about 
fifteen years ago.  This force also leads to very reasonable 
properties of ordinary nuclear matter.  Our theory is therefore 
free of any adjustable parameters.  The energy landscapes 
${\cal {H}}(B,b)$ for various $n\,\alpha$ nuclei are shown in Fig.~\ref{fig:E_surface}~\cite{funaki_be,tohsaki_nara}. We see that they all have qualitatively the same 
structure. From the minimum point on, with increasing $B$-parameter, developes 
a valley with constant $b$ which takes approximately the value of the free 
space $\alpha$-particle. The valley then goes over a saddle point which 
indicates the disintegration of the nucleus into n $\alpha$-particles.
It is interesting to see that the minimum of the energy surfaces does not 
correspond to the Slater determinant case with $b=B$ but rather a quite 
substantial gain in energy due to four body correlations can be observed even 
for the ground state in these light nuclei.

It is evident that for large numbers of $\alpha$-particles the explicit 
antisymmetrisation demanded in our wave function of Eq.~(\ref{eq:2}), will encounter 
great difficulties. On the other hand in the case of many quartets or a 
macroscopic number of it in the condensate, in particular in infinite 
nuclear matter, one may transform our number conserving condensate wave 
function into a coherent state and proceed in a similar way as in the BCS 
case of standard pairing. Roughly this can go along the following lines. 
The first step is to define an $\alpha$-particle coherent state 

\begin{equation}
|\alpha \rangle \sim \exp\Big(\frac{1}{4!}\sum_{1234}\Phi_{1234}^{(\alpha)}c_1^\dagger c_2^\dagger c_3^\dagger c_4^\dagger\Big)|{\rm vac}\rangle .
\end{equation}

Next comes to mix pair creators and pair destructors:
\begin{equation}
Q_{\alpha}^\dagger =\frac{1}{2} \sum_{12}\{ X_{12}^{\alpha}c_1^\dagger c_2^\dagger - Y_{12}^{\alpha}c_1c_2\} .
\end{equation}

If we replaced the pair operators by bosons, this would correspond to a 
Hartree-Fock-Bogoliubov approach for bosons, with, in analogy to the fermion 
case, a bosonic gap equation, etc. However, here we want to keep the fermionic structure of the pairs. The amplitudes $X,Y$ shall obey the 
following orthonormality, respectively completeness relations $XX^\dagger - YY^\dagger = 1$. We also demand: $Q_{\alpha}|\alpha \rangle=0$, and the corresponding order parameter is a quartet expectation value: $\langle c_1^\dagger c_2^\dagger c_3^\dagger c_4^\dagger \rangle \sim XY$.

A self-consistent set of quartet equations can then be constructed:
\begin{equation}
\left(
\begin{array}{cc}
A & \Delta_4\\
-\Delta_4 & -A\\
\end{array}
\right)
\left(
\begin{array}{cc}
X\\
Y\\
\end{array}
\right)
=E
\left(
\begin{array}{cc}X\\
Y\\
\end{array}
\right),
\end{equation}
where
\begin{equation}
\Delta_4 = \langle [c^\dagger c^\dagger ,[H,c^\dagger c^\dagger ]]\rangle \sim v_{....}\langle c_1^\dagger c_2^\dagger c_3^\dagger c_4^\dagger \rangle, 
\end{equation}
and $v_{....}$ stands for the matrix element of the interaction and the bosonic ``gap'' is then very schematically given by
\begin{equation}
\Delta_4 \sim \sum v_{....}XY.
\end{equation}

Similar equations from an analogous procedure can be obtained mixing a fermion destructor with three creators: $q_\alpha^\dagger=\sum U_{123}^\alpha c_1^\dagger c_2^\dagger c_3^\dagger -\sum V_1^\alpha c_1$. The final equation for quartetting is a very intuitive extension of the equation for the BCS pairing order parameter $\langle cc \rangle$~\cite{Ring_Schuck}.
\begin{equation}
(\epsilon_1+\epsilon_2+\epsilon_3+\epsilon_4)K_{1234}^\alpha - \Big[ (1-n_1-n_2)\Delta_{1234}^\alpha + {\rm perms.} \Big] = 4\mu K_{1234}^\alpha , \label{eq:quart}
\end{equation}
with $\epsilon_i$ the self-consistent single particle energies and $\Delta_{1234}^\alpha =\frac{1}{2} \sum v_{123^\prime 4^\prime} K_{3^\prime 4^\prime 34}^\alpha$ where $K_{1234}$ is the order parameter specified below, and $v_{1234}$ is the antisymmetrised matrix element of the interaction. The occupation numbers $n_i  = \langle c_i^\dagger c_i \rangle$ are obtained from the single particle Green's function $G_1 = (\omega-\epsilon_1 - M_1^\omega)^{-1}$ with the mass operator
$M_1^{\omega}=\frac{1}{3!}\sum_{234}|D_{234}^{\alpha}|^2N^0_{234}(\omega+\epsilon_2+\epsilon_3 + \epsilon_4)^{-1}$ with $\Delta_{234}^{\alpha}=N^0_{234}D^{\alpha}_{234}$ and $N^0_{234} = (1-n_2)(1-n_3)(1-n_4) + n_2n_3n_4$ (all energies counted from $\mu$). In the zero density limit Eq.~(\ref{eq:quart}) goes over into the exact free space $\alpha$-particle Schr\"odinger equation.

In infinite matter, the order parameter has zero total momentum and we write
\begin{equation}
K_{1234} = \langle c_1^\dagger c_2^\dagger c_3^\dagger c_4^\dagger \rangle \rightarrow \delta({\vec k}_1+{\vec k}_2 + {\vec k}_3+ {\vec k}_4)
\Big\langle c^\dagger_{{\vec k}_1}c^\dagger_{{\vec k}_2}c^\dagger_{{\vec k}_3}c^\dagger_{{\vec k}_4}\Big\rangle_{S,T},
\end{equation}
where $S,T$ stand for the spin-isospin wave function.
Thus, the order parameter depends on three momenta ${\vec \kappa} = 
{\vec k}_1-{\vec k}_2$; ${\vec \kappa}' = {\vec k}_3-{\vec k}_4$; ${\vec P}={\vec k}_1+{\vec k}_2=-({\vec k}_3+{\vec k}_4)$. 
In general this makes 9 variables! Nonlinear equations for the order parameter have to be solved what is a very 
demanding numerical task.
\begin{figure}[htbp]
\begin{center}
\includegraphics[scale=0.55]{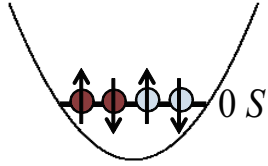}
\vspace*{8pt}
\caption{Spin-isospin saturated $(0S)^4$ mean field configuration of the 
$\alpha$-particle.}\label{fig:0s}
\end{center}
\end{figure}

 However, strong simplifications may be possible! For Bose-Einstein condensation of fermion-clusters heavier than a pair, 
one may proceed to a mean-field description of the cluster with projection on 
good total momentum ${\vec K}$. The clusters condense in the ${\vec K}=0$ 
state. Even for the $\alpha$-particle a mean field description is quite a 
good approximation, under the condition to use effective forces, such as 
Skyrme or Gogny, and to project, as mentioned, on good linear momentum (see Fig.~\ref{fig:0s}).
\begin{equation}
\Phi^{(\alpha)} \sim\delta({\vec k}_1+{\vec k}_2 + {\vec k}_3+ {\vec k}_4)
\varphi_{0S}(\vec{k}_1)\varphi_{0S}(\vec{k}_2)\varphi_{0S}(\vec{k}_3)\varphi_{0S}(\vec{k}_4).
\end{equation}

%In the same way as in the case of pairing we can obtain an equation for the 
%order parameter also for the case of quartetting. Schematically:
%\begin{equation}
%2AXY + (1+2Y^2)\Delta_4 = 0.
%\end{equation}
%
%We see that the phase space factor $(1+2Y^2)$ in front of the quartet ``gap'' has now a bosonic form (fermionic form: $(1-2v^2)$, see eq.~(\ref{eq:1})).

With
\begin{equation}
\Phi^{(\alpha)}_{1234}= (YX^{-1})_{1234},
\end{equation}
and 
\begin{equation}
XX^\dagger - YY^\dagger =1 \rightarrow XX^\dagger=(1-{\Phi^{(\alpha)}}^2)^{-1},
\end{equation}
the self-consistent quartet equation  can be expressed entirely in terms 
of $\Phi^{\alpha}$ and then via the product ansatz everything via a single 
$0S$ wavefunction. The nonlinear equation for $\varphi_{0S}({\vec r})$ should be
solvable! This for any number of $\alpha$-particles! The last part of this section is onging work with T. Sogo~\cite{sogo}.

\section{Results for Finite Nuclei: $^{12}$C}

\begin{figure}[h]
%\begin{center}
\hspace{0.cm}
\begin{minipage}[h]{8cm}
\psfig{figure=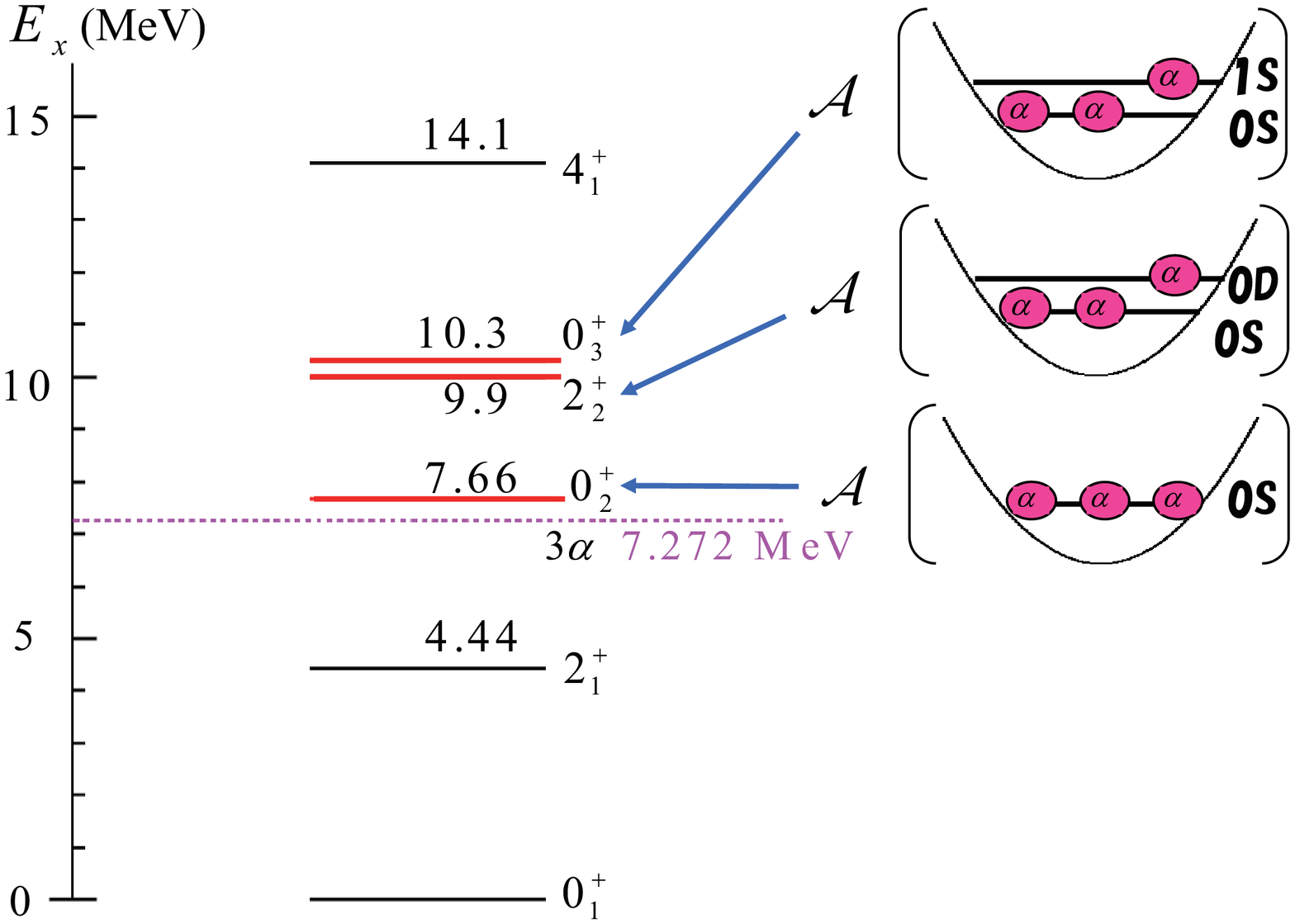,scale=0.43,angle=270}
\caption{(Color online) Theoretical interpretations of the $0_2^+$, $2_2^+$ and $0_3^+$ states, which are missing in the results of no-core shell-model calculation (see Fig.~\ref{fig:exp_12C_nocore}), are schematically shown. }
\end{minipage}
\label{fig:exp_12C}
\hfill
\hspace{-0.4cm}
\begin{minipage}[h]{8.5cm}
\psfig{figure=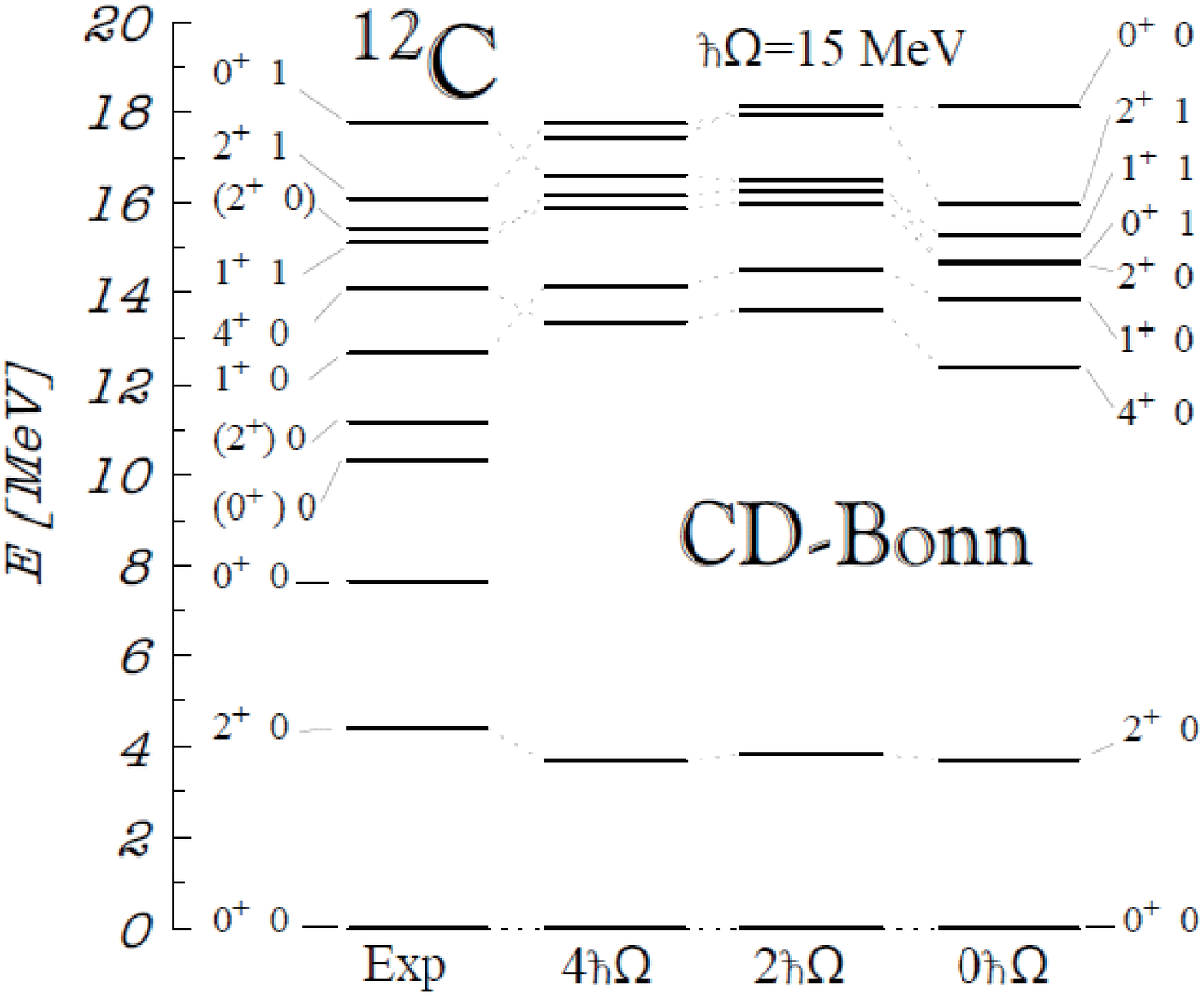,scale=0.35,angle=270}
\caption{Experimental and no core shell model positive-parity excitation spectra of $^{12}$C. The effective interaction was derived from the
CD-Bonn NN potential in a HO basis with $\hbar \Omega= 15$ MeV. Figure taken from \cite{nocore}.}
\label{fig:exp_12C_nocore}
\end{minipage}
%\end{center}
\end{figure}

As we discussed already, the variational
wave function constructed from the Hill-Wheeler equation based on 
Eqs.~(\ref{eq:2}), (\ref{eq:3}), and (\ref{eq:4}) has practically 
100 percent overlap with the RGM and GCM wave functions constructed in 
Refs.~\cite{kamimura} and \cite{uegaki}, once the same 
Volkov force is used \cite{cbec}.  It is, therefore, not astonishing 
that our results are very similar to theirs. Nevertheless, let us again 
discuss the situation in some detail.  For $^{12}$C we obtain 
two eigenvalues in the $0^+$-channel: the ground state and the Hoyle state. 
Theoretical values 
for positions, rms values, and transition probabilities are 
given in Table~\ref{tab:1} and compared to the data. Inspecting
the rms radii, we see that the Hoyle state has a volume 3 to 4 larger 
than that of the ground state of $^{12}$C.  This is the primary 
aspect of the dilute-gas state we highlighted above. 
We also can make a deformed calculation in allowing the width parameter $B$ 
to have different values in the different directions. Projecting on good 
angular momentum then yields the position of the second $2^+$-state in $^{12}$C 
which is in good agreement with the experimental value~\cite{ito,fthsr}. Also its width 
can be evaluated and one obtains a quite reasonable estimate. Detailed 
investigation of the wave function of the $2_2^+$-state shows that it can 
essentially be described in lifting out of the condensate state with the 
three $\alpha$'s in the $0S$-orbit, one $\alpha$-particle in the next 
$0D$-orbit. It is tempting to imagine that the $0_3^+$-state which, 
experimentally, is almost degenerate with the $2_2^+$-state, is obtained by 
lifting one $\alpha$-particle into the $1S$-orbit. Preliminary theoretical 
studies~\cite{kato} indicate that this scenario might indeed apply. However, the 
width of the $0_3^+$ state is very broad ($\sim$ 3 MeV), rendering a 
theoretical treatment rather delicate. Further investigations are necessary 
to validate or reject this picture which is shown graphically in 
Fig.~\ref{fig:exp_12C}. %{\it please give here the experimental spectrum of 12C (eventually together with the shell model results as in Barrett) indicating the triple of states 02,22, and 03, together with the theoretical explanation, lifting the alphas into $0D$ or $1S$ sates, a figure which you give in your talk}. 
 At any rate, it would be 
quite satisfying, if the triplet of states, ($0_2^+,2_2^+, 0_3^+$) could all 
be explained from the $\alpha$-particle perspective, since those three states 
are {\it precisely} the ones which cannot be explained within a (no core) 
shell model approach~\cite{nocore}.

Constructing 
a pure-state $\alpha$-particle density matrix $\rho(\vec{R},\vec{R}')$ 
from our wave function, integrating out of the total density matrix 
all intrinsic $\alpha$-particle coordinates, and diagonalizing this 
reduced density matrix, we find that the corresponding $0S$ 
$\alpha$-particle orbit is occupied to 70 percent by the 
three $\alpha$-particles \cite{yamada1,suzuki}.  This is a huge 
percentage, giving vivid support to the view that the Hoyle
state is an almost ideal $\alpha$-particle condensate. 
\index{alpha cluster!condensate}  We will dwell on this point in more detail 
in sections 7 and 8.

\begin{table}
\begin{center}
\begin{tabular}{ccccc}
\hline\hline
 &  & condensate w.f. & \raisebox{-1.8ex}[0pt][0pt]{RGM \cite{kamimura}} & \raisebox{-1.8ex}[0pt][0pt]{Exp.} \\
 &  & (Hill-Wheeler) &  &  \\
\hline
\raisebox{-1.8ex}[0pt][0pt]{$E$(MeV)} & $0_1^+$ & $-89.52$ & $-89.4$  & $-92.2$  \\
 & $0_2^+$ & $-81.79$ & $-81.7$  & $-84.6$  \\
\hline
\raisebox{-1.8ex}[0pt][0pt]{$R_{\rm r.m.s.}$(fm)} & $0_1^+$ &   $\ \ \ 2.40$ &   $\ \ \ 2.40$ &   $\ \ \ 2.44$ \\
 & $0_2^+$ &   $\ \ \ 3.83$ &   $\ \ \ 3.47$ &  \\
\hline
$M(0_2^+\rightarrow 0_1^+)$(fm$^2$) &  &   $\ \ \ 6.45$ &   $\ \ \ 6.7$ &   $\ \ \ 5.4$  \\
\hline\hline
\end{tabular}
\caption{Comparison of the binding energies, rms radii $(R_{\rm r.m.s.})$, 
and monopole matrix elements $(M(0_2^+\rightarrow 0_1^+))$ for 
$^{12}$C given by solving Hill-Wheeler equation  
  based on Eq.~(\ref{eq:2}) and by Ref.~\cite{kamimura}. 
  The effective two-nucleon force Volkov No.~2 was adopted 
  in the two cases for which the $3\alpha$ threshold energy is
  calculated to be $-82.04$ MeV.}
\label{tab:1}
\end{center}
\end{table}

%%%%%%%%%%%%%%%%%%%%%%%%%%%%%%%%%%%%%%%%%%%%%%%%%%%%%%%%%%%%%%%%%%%%
\begin{figure}[h]
%\begin{center}
\hspace{3.cm}
\begin{minipage}[h]{5.cm}
\psfig{figure=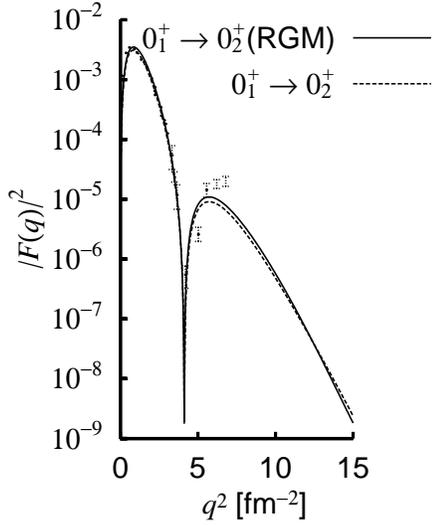,scale=0.8}
\end{minipage}
\hfill
\hspace{-0.4cm}
\begin{minipage}[h]{8.cm}
\caption{Experimental values of inelastic form factor in $^{12}$C to
  the Hoyle state are compared with our values and those given by
  Kamimura et al. in Ref.~\cite{kamimura} (RGM).
  In our result, the Hoyle-state wave function is obtained by
  solving the Hill-Wheeler equation based on Eq.~(\ref{eq:2}).
  The experimental values are taken from Ref.~\cite{fmfct_exp}.}
\label{fig:2_inel}
\end{minipage}
%\end{center}
\end{figure}

\begin{figure}[h]
\hspace{0.5cm}
\begin{minipage}[h]{8.5cm}
\psfig{figure=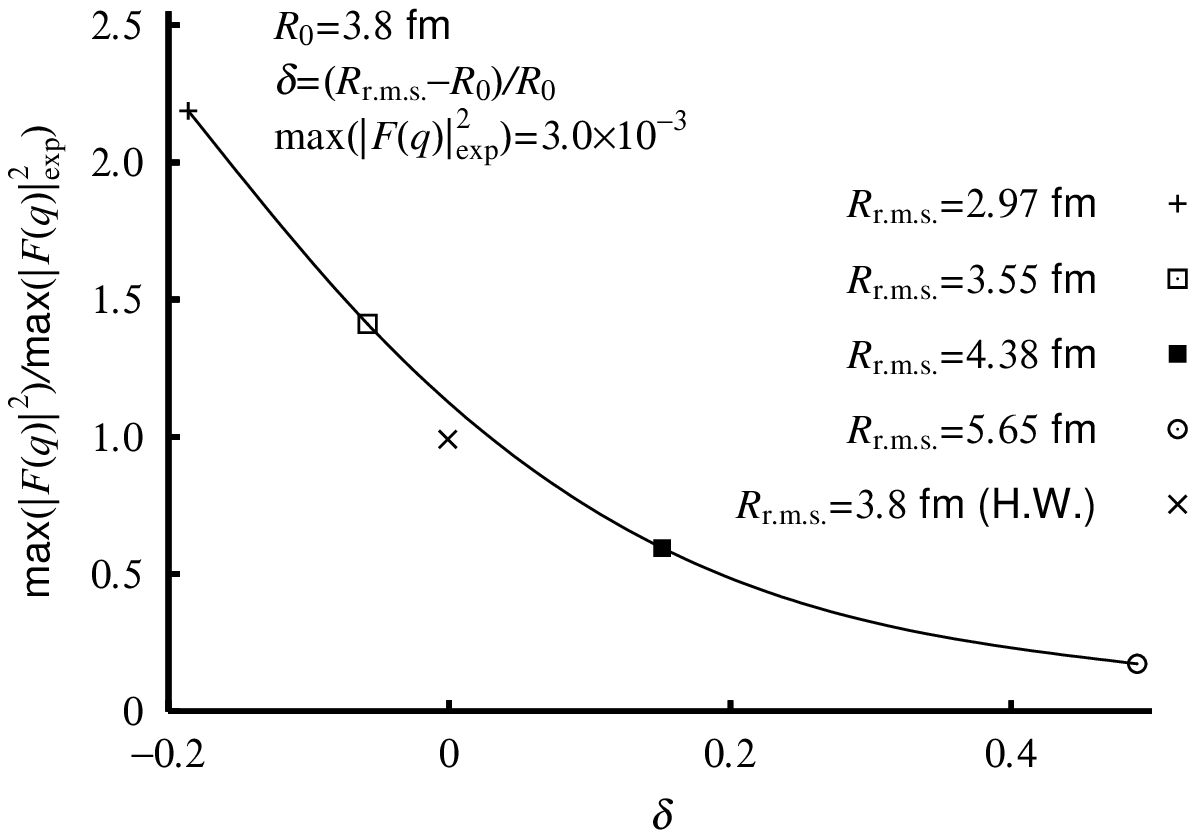,width=8.5cm}
\end{minipage}
\hfill
\hspace{-0.5cm}
\begin{minipage}[h]{7.5cm}
\caption{The ratio of the value of the maximum height, theory versus
  experiment, of the inelastic form factor,
  i.e. max$|F(q)|^2$$/$max$|F(q)|_{\rm exp}^2$, is plotted as a
  function of $\delta=(R_{\rm r.m.s.}-R_0)/R_0$.  Here $R_{\rm r.m.s.}$
  and $R_0$ are the rms radii corresponding, respectively, to the
  wave function of Eq.~(\ref{eq:2}) and that obtained by solving
  the Hill-Wheeler equation based on Eq.~(\ref{eq:2}).}
\label{fig:3_height}
\end{minipage}
\end{figure}
	    
Let us now discuss what to our mind is the most convincing evidence 
that our description of the Hoyle state is the correct one. 
Like the authors of Ref.~\cite{kamimura}, we reproduce very 
accurately the inelastic form factor $0_1^+ \rightarrow 0_2^+$
of $^{12}$C, as shown in Fig.~\ref{fig:2_inel}.  As such, the agreement with 
experiment is already quite impressive.  Additionally, however, the
following study was made, results from which are presented in 
Fig.~\ref{fig:3_height}.  We artificially varied the extension of the 
Hoyle state and examined the influence on the form factor.  It 
was found that the overall shape of the form factor shows 
little variation, for example in the position of the minimum. 
On the other hand, we found a strong dependence of
the absolute magnitude of the form factor; Fig.~\ref{fig:3_height} 
illustrates this behavior with a plot showing the variation of 
the height of the first maximum of the inelastic form factor 
as a function of the percentage change of the rms radius of 
the Hoyle state \cite{funaki1}.  It can be seen that a 20 
percent {\it increase of the rms radius} produces a remarkable
decrease of the maximum  by a factor of 
two! This strong sensitivity of the magnitude of the form 
factor to the size of the Hoyle state enhances our firm 
belief that the agreement with the actual measurement is 
tantamount to a proof that the calculated wide extension 
of the Hoyle state corresponds to reality.

Summarizing our inquiry into the possible role of $\alpha$
clustering in $^{12}$C, we have accumulated enough facts to 
be convinced that the Hoyle state is, indeed, what one may
call an $\alpha$-particle condensate state.  At the same 
time, we acknowledge that referring to only three particles 
as a ``condensate'' constitutes a certain abuse of the word.  
However, in this regard it should be remembered that also
in the case of nuclear Cooper pairing, only a few pairs 
are sufficient to obtain clear signatures of superfluidity 
in nuclei! \\
\indent Let us now go one step further and investigate the four $\alpha$-particle case.

\section{Alpha-particle condensation in $^{16}$O}

 The establishment of this condensate aspect of the Hoyle state naturally 
leads us to the speculation about $4\alpha$-particle condensation in 
$^{16}$O, which is the focus in this section. The situtaion in $^{16}$O is, as 
compared to $^{12}$C quite a bit more complicated, even in the $0^+$ channel 
alone. This stems from the fact that while ``knocking loose'' one $\alpha$-particle in $^{12}$C, necessarily the other two are also almost free ($^8$Be) 
and all three $\alpha$'s form the gas state. However, exciting one $\alpha$-particle out of the ground state in $^{16}$O, may leave the remaining $^{12}$C 
core in the ground state or in various excited states of the shell model type. 
Therefore, in $^{16}$O we need to ``knock loose'' at least two $\alpha$'s to 
obtain the $\alpha$ gas state. 

The $0^+$ spectrum of 
$^{16}$O has, 
in the past, very well been reproduced up to about $13$ MeV excitation energy, 
including 
the ground state, with a semi-microscopic cluster model, i.e. the  
$\alpha + ^{12}$C OCM (Orthogonality Condition Model)~\cite{Suz76}. 
In particular, this model calculation, as well as that of an $\alpha+^{12}$C 
Generator-Coordinate-Method one~\cite{baye2}, demonstrates that the $0_2^+$ 
state at $6.05$ MeV and the $0_3^+$ state at $12.05$ MeV have 
$\alpha + ^{12}$C structures~\cite{Hor68} where the $\alpha$-particle 
orbits around the $^{12}$C$(0_1^+)$-core in an $S$-wave and around the 
$^{12}$C$(2_1^+)$-core in a $D$-wave, respectively. Consistent results were 
later obtained by 
the $4\alpha$ OCM calculation within the harmonic oscillator 
basis~\cite{Kat92}. However, the model space adopted in 
Refs.~\cite{Suz76,baye2,Kat92} is not sufficient to account simultaneously for 
the $\alpha+ ^{12}$C and the $4\alpha$ 
gas-like configurations. On the other hand, the $4\alpha$-particle condensate 
state was first investigated in Ref.~\cite{thsr} and its existence was 
predicted around the $4\alpha$ threshold with the  
$\alpha$-particle condensate wave function. While this so-called THSR 
wave function 
can well describe the dilute $\alpha$ cluster states as well as shell model 
like ground states, other structures such as $\alpha + ^{12}$C clustering 
are smeared out and only incorporated in an average way. Since there exists 
no calculation, so far, which reproduces both the $4\alpha$ gas and 
$\alpha+^{12}$C cluster structures simultaneously, it is crucial to perform 
an extended calculation for the simultaneous reproduction of both kinds of 
structures, which will give a decisive benchmark for the existence 
of the $4\alpha$-particle condensate state from a theoretical point of view.

 Therefore the objective in Ref.~\cite{funaki_4aocm} was to explore the $4\alpha$ condensate 
state by 
solving a full OCM four-body equation of motion without any assumption with 
respect to the structure of the $4\alpha$ system. Here we take the $4\alpha$ 
OCM with Gaussian basis functions, the model space of which is large enough 
to cover the $4\alpha$ gas, the  $\alpha +^{12}$C cluster, as well as the 
shell-model configurations. The OCM is extensively described in 
Ref.~\cite{saitoh}. Many successful applications of OCM are reported in 
Ref.~\cite{carbon}.
The $4\alpha$ OCM Hamiltonian is given as follows:

\begin{equation}
{\cal H}=\sum_{i}^{4}T_i - T_{\rm cm}+ \sum_{i<j}^4
\Big[ V_{2\alpha}^{({\rm N})}(i,j)+V^{({\rm  C})}_{2\alpha}(i,j)
 + V_{2\alpha}^{({\rm P})}(i,j) \Big]  +\sum_{i<j<k}^4 
V_{3\alpha}(i,j,k)+ V_{4\alpha}(1,2,3,4), \label{eq:hamil}
\end{equation}
where $T_i$, $V_{2\alpha}^{({\rm N})}(i,j)$, $V_{2\alpha}^{({\rm C})}(i,j)$, 
$V_{3\alpha}(i,j,k)$ and $V_{4\alpha}(1,2,3,4)$ stand for the operators of 
kinetic energy for the $i$-th $\alpha$ particle, two-body, Coulomb, three-body 
and four-body forces between $\alpha$ particles, respectively. The center-of-
mass kinetic energy $T_{\rm cm}$ is subtracted from the Hamiltonian. 
$V_{2\alpha}^{({\rm P})}(i,j)$ is the Pauli exclusion operator~\cite{kukulin}, 
by which 
the Pauli forbidden states between two $\alpha$-particles in $0S$, $0D$ 
and $1S$ 
states are eliminated, so that the ground state with the shell-model-like 
configuration can be described correctly. The effective $\alpha$-$\alpha$ 
interaction $V_{2\alpha}^{\rm (N)}$ is constructed 
by the folding procedure from two kinds of effective two-nucleon forces. 
One is 
the Modified Hasegawa-Nagata (MHN) force~\cite{mhn} and the other is the 
Schmidt-Wildermuth (SW) force~\cite{sw}, see Refs.~\cite{yamada1} 
and ~\cite{kato} for applications, respectively. We should note 
that the folded $\alpha$-$\alpha$ potentials reproduce the $\alpha$-$\alpha$ 
scattering phase shifts and energies of the $^8$Be ground state and of the 
Hoyle state. The three-body force $V_{3\alpha}$ is as in 
Refs.~\cite{yamada1} and \cite{kato} where it was phenomenologically 
introduced, so as to fit the ground state energy of $^{12}$C. In addition, the 
phenomenological four-body force $V_{4\alpha}$ which is taken to be a Gaussian 
is adjusted to the ground state energy of $^{16}$O, where the range is simply 
chosen to be the same as that of the three-body force. The origin of the three-
body and four-body forces is considered to derive from the state dependence of
 the effective nucleon-nucleon interaction and the additional Pauli repulsion 
between more than two $\alpha$-particles. However, they are short-range, and 
hence only act in compact configurations. The expectation values of those 
forces do not exceed 7 percent of the one of the corresponding two-body term, 
even for the ground state with the most compact structure, i.e. being the most 
sensitive to those forces. 

 Employing the Gaussian expansion method~\cite{GEM} for the choice of 
variational basis functions, the total wave function $\Psi$ of the $4\alpha$ 
system is expanded in terms of Gaussian basis functions as follows: 
\begin{eqnarray}
&&\hspace{-0.7cm} \Psi(0_n^+)=\sum_{c, \nu} A^n_{c}(\nu)\Phi_{c}(\nu), \\
&&\hspace{-0.7cm} \Phi_{c}(\nu) ={\cal \widehat S} \Big[[\varphi_{l_1}
(\vec{r}_1,\nu_1)\varphi_{l_2}(\vec{r}_2,\nu_2)]_{l_{12}}  \varphi_{l_3}
(\vec{r}_3,\nu_3) \Big]_{J}, \label{eq:30}
\end{eqnarray}
where $\vec{r}_1$, $\vec{r}_2$ and $\vec{r}_3$ are the Jacobi coordinates 
describing internal motions of the $4\alpha$ system. ${\cal \widehat S}$ 
stands for the symmetrization operator acting on all $\alpha$ particles 
obeying Bose statistics. $\nu$ denotes the set of size parameters $\nu_1,
\nu_2$ and $\nu_3$ of the normalized Gaussian function, $\varphi_{l}(\vec{r},
\nu_i)=
N_{l,\nu_i}r^l\exp{(-\nu_i r^2)} Y_{l m}(\hat{\vec r})$, and $c$ the set of 
relative orbital angular momentum channels $[[l_1,l_2]_{l_{12}},l_3]_J$ 
depending on either of  the coordinate type of $K$ or $H$~\cite{GEM}, 
where $l_1$, $l_2$ and $l_3$ are the orbital angular momenta with respect 
to the corresponding Jacobi coordinates. The coefficients $A^n_{c}(\nu)$ 
are determined according to the Rayleigh-Ritz variational principle. 
\begin{figure}[htbp]
\begin{minipage}{8cm}
\includegraphics[width=7.1cm]{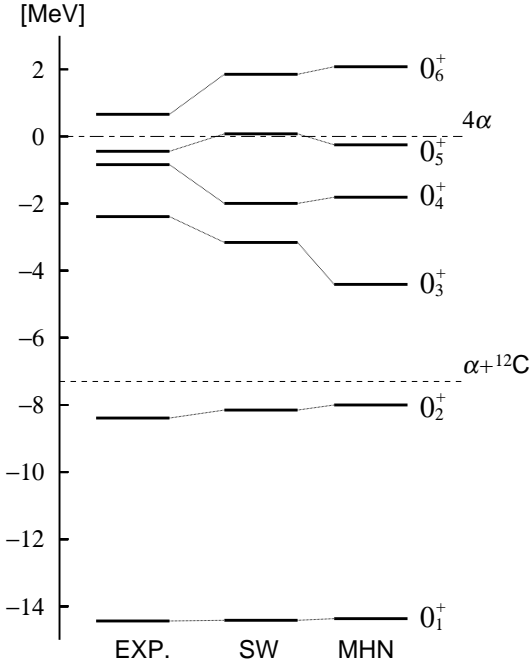}
\end{minipage}
\begin{minipage}{9cm}
\caption{Comparison of energy spectra between experiment and the present 
calculation. Two kinds of effective two-body nucleon-nucleon forces MHN and SW 
are adopted (see text). Dotted and dash-dotted lines denote the 
$\alpha + ^{12}$C and $4\alpha$ thresholds, respectively. Experimental data 
are taken from Ref.~\cite{ajze}, and from Ref.~\cite{wakasa} for the $0_4^+$ 
state. The assignments with experiment are tentative, see, however, detailed 
discussion in the text.}\label{fig:1}
\end{minipage}
\vspace{-0.5cm}
\end{figure}

 Figure~\ref{fig:1} shows the energy spectrum with $J^\pi=0^+$, which is 
obtained by diagonalizing the Hamiltonian, Eq.~(\ref{eq:hamil}), in a model 
space as large as given by 5120 Gaussian basis functions, Eq.~(\ref{eq:30}) 
(the other multipolarities, needing larger basis sets, are more difficult and 
shall be studied in future work). It is confirmed that all levels are well 
converged. With the above mentioned effective $\alpha$-$\alpha$ forces, we 
can reproduce the 
full spectrum of $0^+$ states, and tentatively make a one-to-one 
correspondence of those states with the six lowest $0^+$ states of the 
experimental spectrum. In view of the complexity of the situation, the 
agreement is considered to be very satisfactory. 
\begin{table}
\begin{center}
\caption{The rms radii $R$ and monopole transition matrix elements to the 
ground state $M({\rm E}0)$ in units of fm and fm$^2$, respectively. 
$R_{\rm exp.}$ and $M({\rm E}0)_{\rm exp.}$ are the corresponding experimental 
data. The finite-size effect of $\alpha$ particle is taken into account in $R$ 
and $M({\rm E}0)$ (see Ref.~\cite{yamada1} for details).}\label{tab:2}
\begin{tabular}{cccccccccc}
\hline\hline
 & & \multicolumn{2}{c}{$R$} & & \multicolumn{2}{c}{$M({\rm E}0)$} & & $R_{\rm exp.}$ & $M({\rm E}0)_{\rm exp.}$  \\
\hline
 & & SW & MHN & & SW & MHN & & & \\
\hline
$0_1^+$ & & $2.7$ & $2.7$ & &        &       & & $2.71\pm0.02$ &          \\
$0_2^+$ & & $3.0$ & $3.0$ & &  $4.1$ & $3.9$ & &        &  $3.55\pm 0.21$  \\
$0_3^+$ & & $2.9$ & $3.1$ & &  $2.6$ & $2.4$ & &        &  $4.03\pm 0.09$  \\
$0_4^+$ & & $4.0$ & $4.0$ & &  $3.0$ & $2.4$ & &        &  no data \\
$0_5^+$ & & $3.1$ & $3.1$ & &  $3.0$ & $2.6$ & &        &  $3.3\pm0.7$   \\
$0_6^+$ & & $5.0$ & $5.6$ & &  $0.5$ & $1.0$ & &        &  no data \\
\hline\hline
\end{tabular}
\end{center}
\vspace{-0.5cm}
\end{table}
We show in Table~\ref{tab:2} the calculated rms radii and monopole matrix 
elements to the ground state, together with the corresponding experimental 
values.
The $M({\rm E}0)$ values for the $0_2^+$ and $0_5^+$ states are 
consistent with the corresponding experimental values. The consistency for 
the $0_3^+$ state is within a factor of two. As mentioned above, the 
structures of the $0_2^+$ and $0_3^+$ states are well established as having 
the $\alpha + ^{12}$C$(0_1^+)$ and $\alpha + ^{12}$C$(2_1^+)$ 
cluster structures, respectively. These structures of the $0_2^+$ and $0_3^+$ 
states are confirmed in the present calculation. We also mention that the 
ground state is described as having a shell-model configuration within the 
present framework, the calculated rms value agreeing with the observed 
one ($2.71$ fm). 

On the contrary, the structures of the observed $0_4^+$, $0_5^+$ and $0_6^+$ 
states in Fig.~\ref{fig:1} have, in the past, not clearly been understood, 
since they have never been discussed with the previous cluster model 
calculations~\cite{Suz76,baye2,Kat92}. Although Ref.~\cite{thsr}, using the THSR wave function, predicts the 
$4\alpha$ condensate state around the $4\alpha$ threshold, it is not clear to 
which of those states it corresponds to. We will analyse the situation with 
the THSR wave function of \cite{thsr} in a future publication~\cite{thsr_prepare}.

 As shown in Fig.~\ref{fig:1}, the present calculation succeeded, for the 
first time, to reproduce the $0_4^+$, $0_5^+$ and $0_6^+$ states, together 
with the $0_1^+$, $0_2^+$ and $0_3^+$ states. This puts us in a favorable 
position to discuss the $4\alpha$ condensate state, expected to exist around 
the $4\alpha$ threshold.

In Table~\ref{tab:2}, the largest rms 
value of about 5 fm is found for the $0_6^+$ state. Compared with the 
relatively smaller rms 
radii of the $0_4^+$ and $0_5^+$ states, this large size suggests that 
the $0_6^+$ state may be composed of a weakly interacting gas of $\alpha$ 
particles~\cite{foot} of the condensate type.
\begin{figure}[htbp]
\begin{minipage}{9cm}
\includegraphics[scale=0.7]{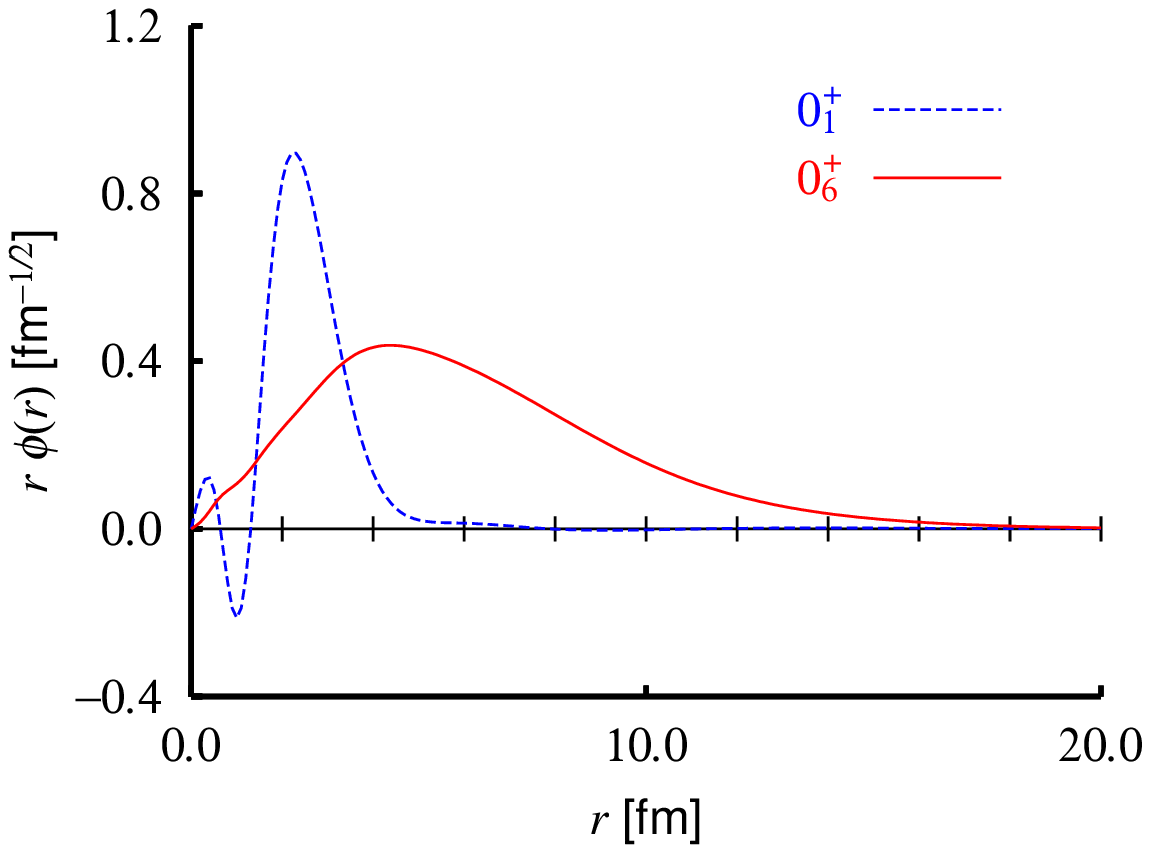}
\caption{(Color online) The radial parts of single-$\alpha$ orbits with $L=0$ 
belonging to the largest occupation number, for the ground and $0_6^+$ states 
with MHN force.}\label{fig:2_occup}
\end{minipage}
\hfill
\begin{minipage}{9cm}
\includegraphics[scale=0.7]{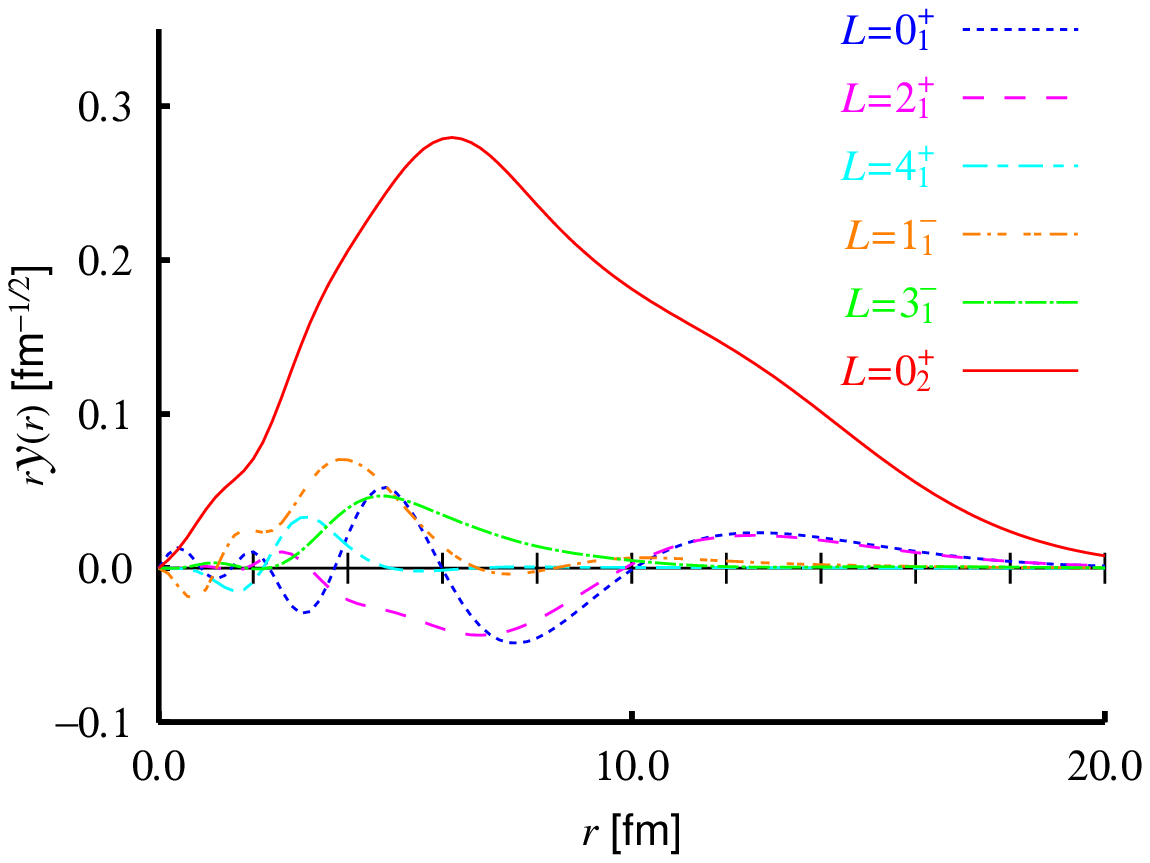}
\caption{(Color online) $r{\cal Y}(r)$ defined by Eq.~(\ref{eq:rwa}) for 
the $0_6^+$ state with the MHN force.}\label{fig:3_red}
\end{minipage}
\end{figure}

While a large size is generally necessary for forming an $\alpha$ 
condensate, the best way for its identification is to investigate the 
single-$\alpha$ orbit and its occupation 
probability, which can be obtained by diagonalizing the one-body ($\alpha$) 
density matrix as defined in ~\cite{takahashi,suzuki,yamada1,density_matrix}.
As a result of the calculation of the $L=0$ case, a large occupation 
probability of $61 \%$ of the lowest $0S$-orbit is found for the $0_6^+$ 
state, whereas 
the other five $0^+$ states all have appreciably smaller values, at 
most $25 \%$ ($0^+_2$). The corresponding single-$\alpha$ $S$ orbit is 
shown in Fig. \ref{fig:2_occup}. It has a strong spatially extended behaviour 
without any node $(0S)$. This indicates that $\alpha$ particles are 
condensed into the very dilute $0S$ single-$\alpha$ orbit, see 
also Ref.~\cite{ropke2}. Thus, the $0^+_6$ state clearly has $4\alpha$ 
condensate character. We should 
note that the orbit is very similar to the single-$\alpha$ orbit of the Hoyle 
state~\cite{suzuki,yamada1}. We also show in Fig.~\ref{fig:2_occup} the 
single-$\alpha$ orbit for the ground state. It has maximum amplitude at 
around $3$ fm and  oscillations in the interior with two nodal $(2S)$ 
behaviour, due to the Pauli principle and reflecting the shell-model 
configuration. 

 In order to further analyze the obtained wave functions, we calculate an 
overlap amplitude, which is defined as follows:
\begin{equation}
{\cal Y}(r)= \Big\langle \Big[ \frac{\delta(r^\prime-r)}{r^{\prime 2}}
Y_{L}(\vec{\hat r}^\prime)\Phi_{L}(^{12}{\rm C}) \Big]_{0} \Big| \Psi(0_6^+) 
\Big\rangle. \label{eq:rwa}
\end{equation}
Here, $\Phi_{L}(^{12}{\rm C})$ is the wave function of $^{12}$C, given 
by the $3\alpha$ OCM calculation~\cite{yamada1}, and $r$ is the relative 
distance between the c.o.m. of $^{12}$C and the $\alpha$ particle. 
From this quantity we can see how large is the component in a certain 
$\alpha + ^{12}$C channel which is contained in our wave function (\ref{eq:30}) 
for $0_6^+$. The 
amplitudes for the $0_6^+$ state are shown in Fig.~\ref{fig:3_red}. It only has 
a large amplitude in the $\alpha + ^{12}$C$(0_2^+)$ channel, whereas 
the amplitudes in other channels are much suppressed. The amplitude in the 
Hoyle-state channel has no oscillations and a long tail stretches out 
to $\sim 20$ fm. This behaviour is very similar to that of the single-$\alpha$ 
orbit of the $0_6^+$ state discussed above. 

 The $\alpha$ decay width constitutes a very important information to 
identify the $0_6^+$ state from the experimental point of view. It can be 
estimated, based on the $R$-matrix theory, with the overlap amplitude 
Eq.~(\ref{eq:rwa})~\cite{r-matrix}. We find that the total $\alpha$ decay 
width of the $0_6^+$ state is as small as 50 keV (experimental value: 
166 keV). This means that the state can be observed as a quasi-stable state. 
Thus, the width, as well as the excitation energy, are consistent with the 
observed data. All the characteristics found from our OCM calculation, 
therefore, 
indicate that the calculated 6th $0^+$ state with 4 alpha condensate 
nature can probably be identified with the experimental $0_6^+$ state 
at $15.1$ MeV.

 Finally we discuss the structures of the $0_4^+$ and $0_5^+$ states. Our 
present calculations show that the $0_4^+$ and $0_5^+$ states mainly have 
$\alpha + ^{12}$C$(0_1^+)$ structure with higher nodal behaviour and 
$\alpha + ^{12}$C$(1^-)$ structure, respectively. Further details will be 
given in forthcoming work. The calculated width of the $0_4^+$ 
is $\sim 150$ keV, which is quite a bit larger than that found for the $0_5^+$ 
state $\sim 50$ keV. Both are qualitatively consistent with the corresponding 
experimental data, $600$ keV and $185$ keV, respectively. The reason why the 
width of the $0_4^+$ state is larger than that of the $0_5^+$ state, though 
the $0_4^+$ state has lower excitation energy, is due to the fact that the 
former has a much larger component of the $\alpha+ ^{12}$C$(0_1^+)$ decay 
channel, reflecting the characteristic structure of the $0_4^+$ state. 
The $4\alpha$ condensate state, thus, should not be assigned to the $0_4^+$ 
or $0_5^+$ state~\cite{4athsr} but very likely to the $0_6^+$ state. 

 In conclusion, the investigation of the $0^+$-spectrum with the 
$4\alpha$ OCM calculation 
succeeded in describing the structure of the full observed $0^+$ spectrum up 
to the $0^+_6$ state in $^{16}$O. The $0^+$ spectrum of $^{16}$O up to about 
15 MeV is now essentially understood, including the $4\alpha$ condensate 
state. This is remarkable improvement concerning our knowledge of the 
structure of $^{16}$O. We find that the $0_6^+$ state above the $4\alpha$ 
threshold has a very large rms radius of about $5$ fm and has a rather large 
occupation probability of $61 \%$ of four $\alpha$ particles sitting in a 
spatially extended single-$\alpha$ $0S$ orbit. The wave function has a large $\alpha + ^{12}$C amplitude only for $^{12}$C$^\ast$, i.e. the 
Hoyle state. These results are strong evidence of the 
$0_6^+$ state, which is a new theoretical prediction, for being the $4\alpha$ 
condensate state, i.e. the analog to the Hoyle state in $^{12}$C. 
Further experimental information is very much requested to confirm the 
novel interpretation of this state. Also independent 
theoretical calculations are strongly needed for confirmation of our results.

\begin{figure}[h]
\hspace{0.7cm}
\psfig{figure=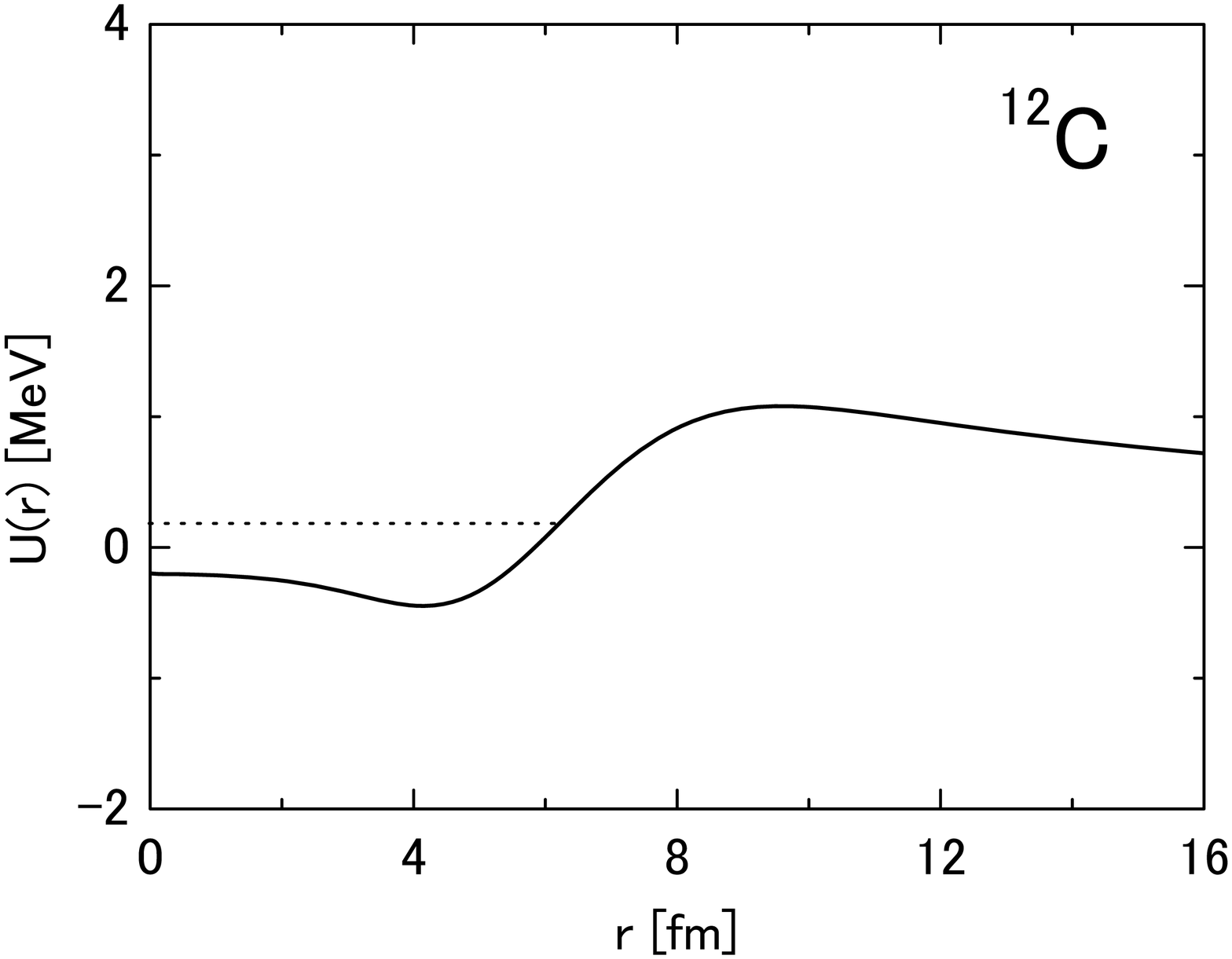,width=8.cm}
\hfill
\hspace{-0.7cm}
\psfig{figure=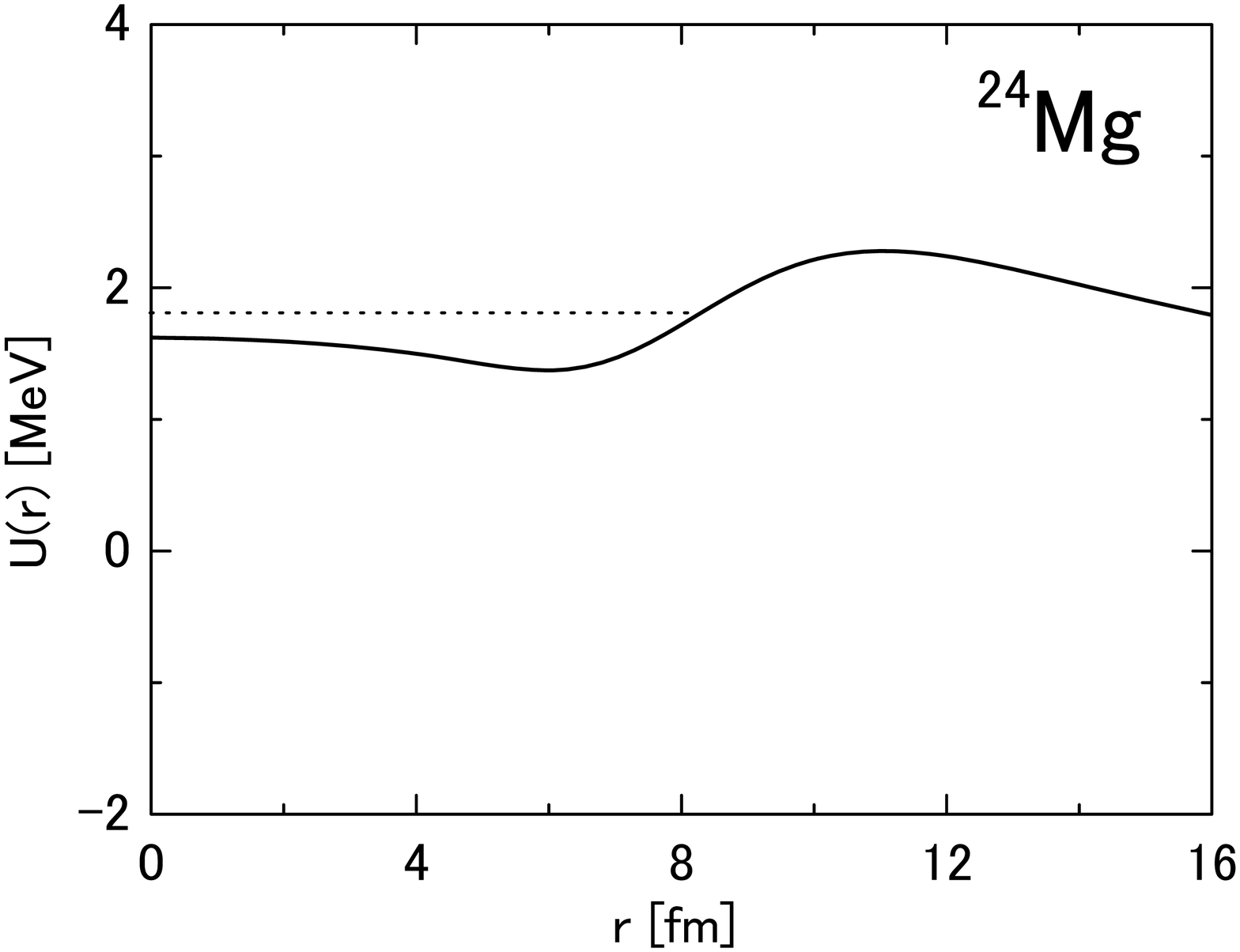,width=8.cm}
\caption{Alpha-particle mean-field potential for three $\alpha$'s
  in $^{12}$C and six $\alpha$'s in $^{24}$Mg. Note the lower Coulomb
  barrier for $^{24}$Mg (from Ref.~\cite{yamada2}).}
\label{fig:4}
\end{figure}

In principle, one could go on, inreasing the number of $\alpha$-particles: 
$^{20}$Ne, $^{24}$Mg, etc. However, one easily imagines that the complexity of the 
calculations quickly becomes prohibitive. In order to get a rough idea what 
happens for more $\alpha$-particles, drastic approximations have to be 
performed. One such approximation is to consider the $\alpha$-particles as 
ideal inert bosons and to treat them in mean field approximation. This then 
leads to the Gross-Pitaevskii Equation (GPE) which is widely employed in the 
physics of cold atoms~\cite{stringari}.
One interesting question that can be asked in this connection is: 
How many $\alpha$'s can maximally exist in a self-bound 
$\alpha$-gas state?  Seeking an answer,
we performed a schematic investigation using an effective 
$\alpha$-$\alpha$ interaction within an $\alpha$-gas mean-field 
calculation of the Gross-Pitaevskii type \cite{gross}. 
The parameters of the force were adjusted to reproduce 
our microscopic results for $^{12}$C. The corresponding 
$\alpha$ mean-field potential is shown in Fig.~\ref{fig:4}. 
One sees the $0S$-state lying slightly above threshold 
but below the Coulomb barrier.  As more $\alpha$-particles
are added, the Coulomb repulsion drives the loosely bound 
system of $\alpha$-particles farther and farther apart, so that the 
Coulomb barrier fades away.  According to our estimate \cite{yamada2}, 
a maximum of eight to ten $\alpha$-particles can be held together in a 
condensate.  However, there may be ways to lend additional stability
to such systems.  We know that in the case of $^8$Be, adding one or two 
neutrons produces extra binding without seriously disturbing the pronounced 
$\alpha$-cluster structure.  Therefore, one has reason to speculate 
that adding a few of neutrons to a many-$\alpha$ state may 
stabilize the condensate.  But again, state-of-the-art microscopic
investigations are necessary before anything definite can be said 
about how extra neutrons will influence an $\alpha$-particle 
condensate. A study in this direction is given in Ref.~\cite{Itagaki} for $^{14}$C.

\section{Alpha-particle condensates in finite nuclei and the alpha-particle 
occupation numbers}

\begin{figure}[h]
\hspace{0.4cm}
\begin{minipage}[h]{8cm}
\psfig{figure=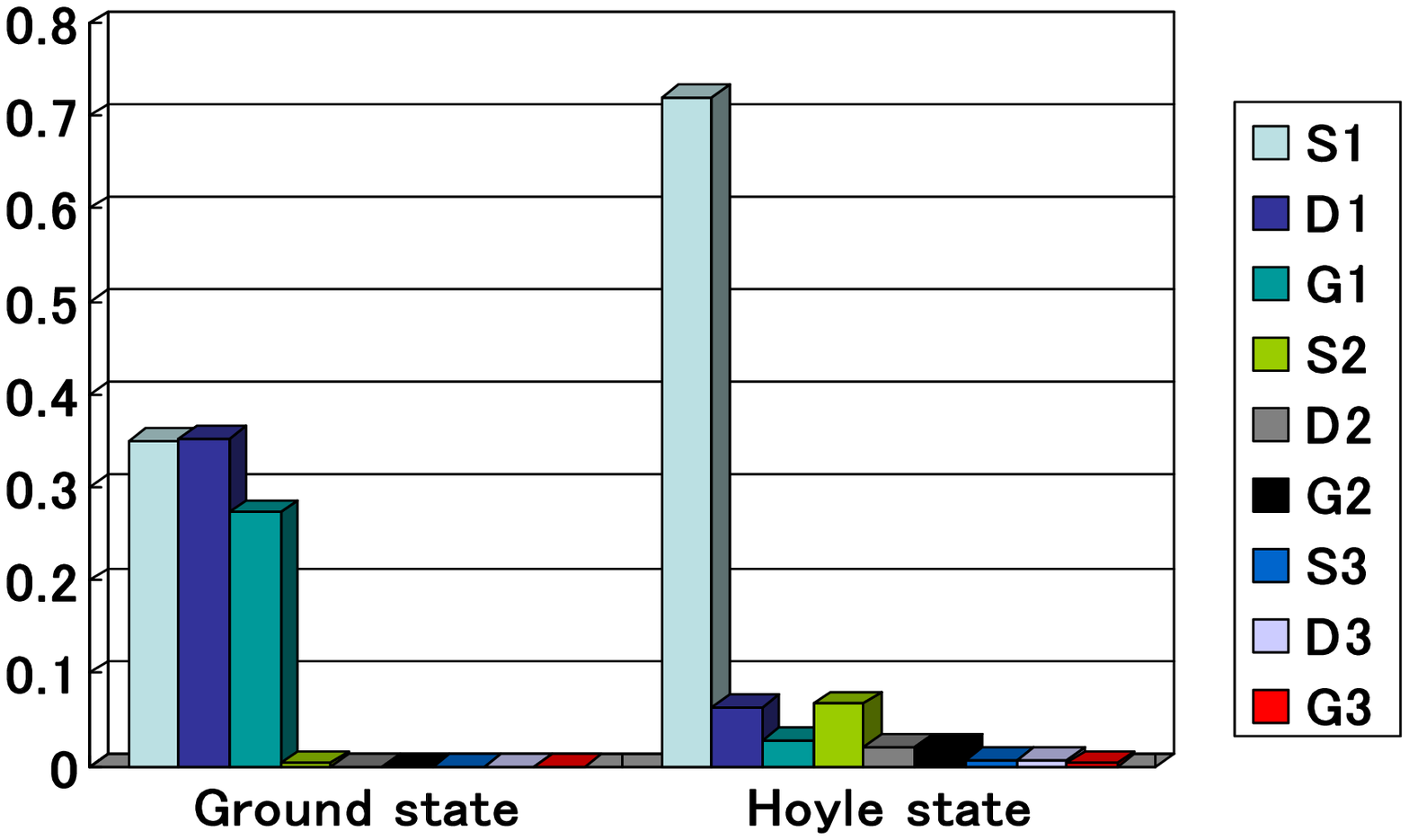,scale=0.5}
\caption{(Color online) 
Occupation of the single-$\alpha$ orbitals of the ground state of $^{12}$C compared with the Hoyle state~\cite{yamada1}.}\label{fig:hist}
\end{minipage}
\hfill
\hspace{-1cm}
\begin{minipage}[h]{9.5cm}
\psfig{figure=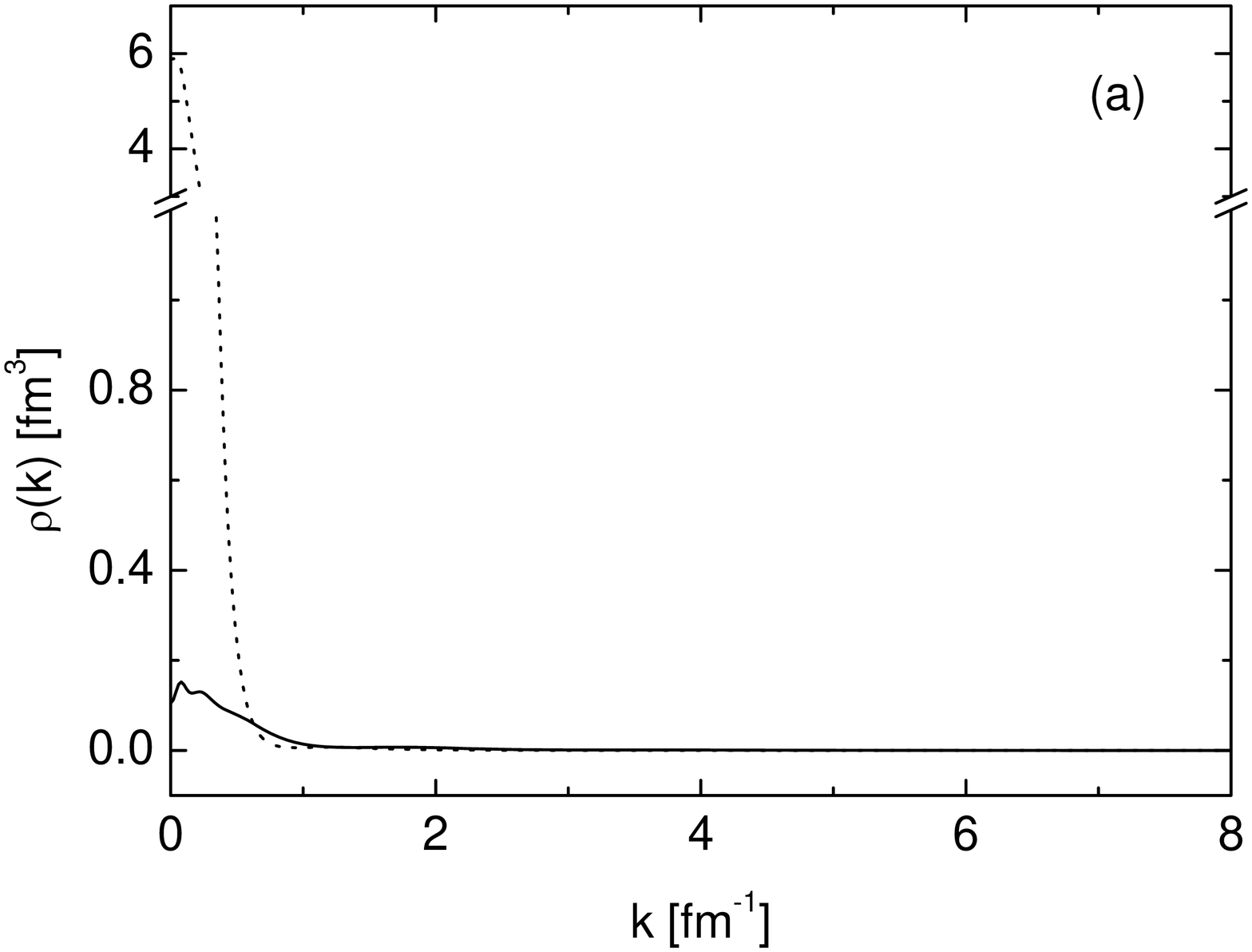,scale=0.35}
\caption{Momentum distribution of the $\alpha$ particle for the ground (solid line) and $0_2^+$ (dotted line) states.}\label{fig:mom}
\end{minipage}
\end{figure}

As already mentioned, constructing an $\alpha$-particle density matrix 
$\rho({\vec R}, {\vec R}')$ by integrating out of the total density matrix all 
intrinsic $\alpha$-particle coordinates and diagonalising the result, one finds 
that the corresponding $0S$ $\alpha$-particle orbit in $^{12}$C is occupied to more 
than 70 percent by the three $\alpha$-particles~\cite{suzuki,yamada1}. This is a huge percentage, affirming the almost ideal $\alpha$-particle condensate nature of the Hoyle state. By contrast, even at zero 
temperature only ten percent of the particles in superfluid $^4$He belong to the 
condensate (which is nevertheless a macroscopic supply of condensed 
particles). To add further perspective to the picture, in the {\it ground state}  
of $^{12}$C the $\alpha$-particle occupations are equally shared between $S,D$ 
and $G$ orbits, thus invalidating a condensate picture for the ground state. 
The occupation numbers for the ground and Hoyle states are shown in histogramm 
format in Fig.~\ref{fig:hist}. The difference between the Hoyle state and the ground 
state is seen to be spectacular. In the Hoyle state the $0S$-occupancy is at 
least an order of magnitude(!) higher than for any other orbit. This is one 
of the main typical features of Bose-Einstein condensation, even in strongly 
correlated Bose systems where there may be a strong depletion of the 
condensate, like in superfluid $^4$He. On the other hand, the ground state 
occupancies can be explained quite well with the standard shell model~\cite{yamada1}. It should also be noted that the ground state 
of $^{12}$C is reasonably well reproduced by our theory (see Table~\ref{tab:1}). 
A further strong indication of the condensate-like behavior of the $\alpha$-particles in the Hoyle state is their momentum distribution, which is much 
narrower, almost delta-function-like, than in the ground state, see Fig.~\ref{fig:mom}. On the same line of investigation one finds, as already mentioned in section 6, that the 6-th $0^+$ state in $^{16}$O has 61\% $0S$ occupancy of 
$\alpha$-particles. 

\begin{figure}[htbp]
\begin{center}
\includegraphics[scale=0.7]{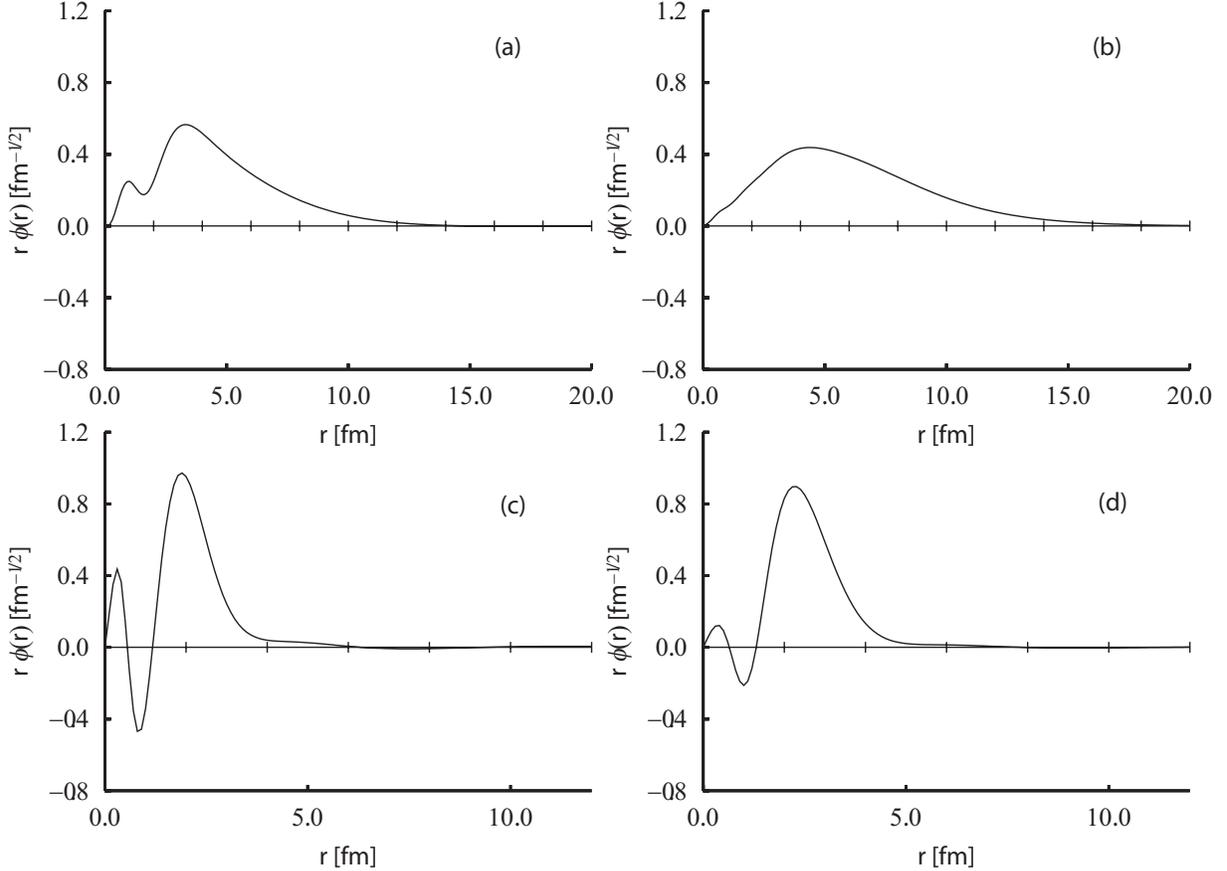}
\caption{Radial parts of the single-$\alpha$ $S$ orbits, (a) of the Hoyle state ($^{12}$C), (b) of the $0_6^+$ state in $^{16}$O and of the ground states (c) in $^{12}$C and (d) in $^{16}$O.}\label{fig:1_via}
\end{center}
\end{figure}

In Fig.~\ref{fig:1_via}(a) and (b), we show radial parts of the single-$\alpha$ $S$ orbits of the Hoyle state~\cite{yamada1} and the $0_6^+$ state in $^{16}$O~\cite{funaki_4aocm}, respectively. 
We see an almost identical shape! 
Of course, the extension is slightly different because of the smallness of the system.
The nodeless character of the wave function is very pronounced and only some oscillations
 with small amplitude are present in $^{12}$C, reflecting a weak influence of the Pauli principle between the $\alpha$'s! 
On the contrary, we show in Fig.~\ref{fig:1_via}(c) and (d), radial parts of the single-$\alpha$ $S$ orbits of the ground states
 in $^{12}$C~\cite{yamada1} and $^{16}$O~\cite{funaki_4aocm}, respectively. 
Due to its much reduced radius the ``$\alpha$-like'' clusters strongly overlap, producing strong amplitude
 oscillations which take care of antisymmetrisation between clusters. 
Again this example very impressively demonstrates the condensate nature of the Hoyle state and
 the $0_6^+$ state in $^{16}$O.
%This result is much in contrast with the fact that ZJ announced the similarity criterion for $\alpha$-particles being very difficult to be fulfilled in finite systems with only a few bosons.

 It is worth noting at this point that the definition of the $\alpha$-particle 
density matrix, that is of a self bound system, is somewhat ambiguous and that 
different definitions may lead to different answers for the occupancies. This 
question has recently been debated in a number of papers~\cite{jensen,funaki4,density_matrix}. Our conclusion 
concerning this point is that one should use for the definition of the 
internal density matrix an orthogonal systems of independent coordinates, like 
they are for instance given by the Jacobi coordinates. This definition insures 
the physically very reasonable boundary condition that, given a system 
exhibits an ideal Bose condensate in the laboratory system, it so remains in 
the internal system and both description become equivalent in the 
thermodynamic limit~\cite{density_matrix}. Jacobi coordinates have been used to evaluate the 
occupancies of $\alpha$-particles in nuclei mentioned above~\cite{yamada1,suzuki,funaki_4aocm,takahashi}.

\section{Reduction of the $\alpha$-condensate with
 increasing density}

\index{alpha cluster!condensate}
The properties of $\alpha$ matter can be used to frame the 
discussion of the structure of $n\,\alpha$ nuclei.  As described 
in the preceding section, computational studies of these 
nuclei based on THSR cluster states have demonstrated that 
an $\alpha$ condensate is established 
at low nucleon density.  More specifically,
states lying  near the threshold for 
decomposition into $\alpha$ particles, notably the ground
state of $^8$Be, $^{12}$C in the $0_2^+$ Hoyle state, and 
 corresponding states in $^{16}$O and other $n \alpha$ 
nuclei are {\it dilute}, being of low mean density and unusually 
extended for their mass numbers.  We have shown quantitatively within
a variational approach that $\alpha$-like clusters are well formed,
with the pair correlation function of $\alpha$-like clusters predicting
relatively large mean distances.  For example, in determining the
sizes of the $^{12}$C nucleus in its $0_1^+$ (ground) state and in 
its $0_2^+$ excited state, we obtained rms radii of 2.44 fm and 
3.83 fm, respectively.  The corresponding mean nucleon densities 
estimated from $36/4 \pi r^3_{\rm rms}$ are close to the 
nuclear-matter saturation density $\rho_0= 0.16$ nucleon/fm$^3$ 
in the former state and 0.03 nucleon/fm$^3$ in the latter.
The expected low densities of putative alpha-condensate states 
are confirmed by experimental measurements of form 
factors \cite{funaki1}.

All of our considerations indicate that quartetting is possible 
in the low-density regime of nucleonic matter, and that 
$\alpha$ condensates can survive until densities of about 
0.03 nucleons/fm$^3$ are reached.  Here, we are in the region
where the concept of $\alpha$ matter can reasonably be 
applied \cite{JC80,SMS06}. 
It is then clearly of interest to use this model to gain further
insights into the formation of the condensate, and especially
the reduction or suppression of the condensate due to repulsive 
interactions.  We will show explicitly that in the model of
$\alpha$ matter, as in our studies of finite nuclei, 
condensate formation is diminished with increasing density.  
Already within an $\alpha$-matter model based on a simple 
$\alpha - \alpha$ interaction, we can demonstrate that the 
condensate fraction -- the fraction of particles in the 
condensate -- is significantly reduced from unity at a density of 
0.03 nucleon/fm$^3$ and essentially disappears approaching nuclear 
matter-saturation density.

The quantum condensate formed by a homogeneous interacting boson 
system at zero temperature has been investigated in
the classic 1956 paper of Penrose and Onsager \cite{PO} 
who characterize the phenomenon in terms of off-diagonal 
long-range order of the density matrix.  Here we recall 
some of their results that are most relevant to our problem.  
Asymptotically, i.e., for $|{\vec r} - {\vec r}^\prime | \sim \infty$, 
the nondiagonal density matrix in coordinate representation 
can be decomposed as
\begin{equation}
\rho ({\vec r}, {\vec r}') \sim \psi_0^* ({\vec r}) \psi_0 ({\vec r}')
+ \gamma ({\vec r}- {\vec r}')\,.
\end{equation}
In the limit, the second contribution on the right vanishes, and 
the first approaches the condensate fraction, formally defined by
\begin{equation}
\rho_0 = \frac{\langle \Psi |a_0^\dagger a_0^{} | \Psi \rangle }{ \langle
    \Psi | \Psi \rangle } \,.
\end{equation}
Penrose and Onsager showed that in the case of a hard-core repulsion,
the condensate fraction is determined by a filling factor describing
the ratio of the volume occupied by the hard spheres.
They applied the theory to liquid $^4$He, and found that for
a hard-sphere model of the atom-atom interaction yielding 
a filling factor of about 28\%, the condensate fraction
at zero temperature is reduced from unity (its value for the 
noninteracting system) to around 8\%.  (Remarkably, but to some extent 
fortuitously, this estimate is in rather good agreement with
current experimental and theoretical values for the condensate
fraction in liquid $^4$He.)

To make a similar estimate of the condensate fraction 
for $\alpha$ matter, we follow Ref.~\cite{ST} and 
assume an ``excluded volume'' for $\alpha$ particles of  
20 fm$^3$.  At a nucleonic density of $\rho_0/3$, this
corresponds to a filling factor of about 28\%, the same
as for liquid $^4$He.  Thus, a substantial reduction of 
the condensate fraction from unity (for a noninteracting
$\alpha$-particle gas at zero temperature) is also 
expected in low-density $\alpha$ matter.

Turning to a more systematic treatment, we proceed in
much the same way as Clark and coworkers \cite{JC80}, 
referring especially to the most recent study with 
M. T. Johnson.  Adopting the $\alpha-\alpha$ interaction 
potential 
\begin{equation}
V_{\alpha}(r) = 475\,\, e^{-(0.7 r/{\rm fm})^2} {\rm MeV} - 130\,\,
e^{-(0.475 r/{\rm fm})^2}{\rm MeV} 
\label{AliBodmer}
\end{equation}
introduced by Ali and Bodmer \cite{AB66},
we calculate the reduction of the condensate fraction as function of
density within what is now a rather standard variational
approach.  Alpha matter is described as an extended, uniform
Bose system of interacting $\alpha$ particles, {\it disregarding} 
any change of the internal structure of the $\alpha$ clusters 
with increasing density. In particular, the dissolution of 
bound states associated with Pauli blocking (Mott effect) 
is not taken into account in the present description. 

The simplest form of trial wave function incorporating the 
strong spatial correlations implied by the interaction 
potential (\ref{AliBodmer}) is the familiar Jastrow choice, 
\begin{equation}
\Psi(\vec r_1, \dots, \vec r_A) = \prod_{i<j} 
                                  f(|\vec r_i - \vec r_j|)\,.
\end{equation}
The normalization condition 
\begin{equation}
4 \pi \rho_\alpha \int_0^\infty [f^2(r) - 1]\,\, r^2 dr = -1\,,
\label{norm}
\end{equation}
in which $\rho_{\alpha}$ is the number density of $\alpha$-particles,
is imposed as a constraint on the variational wave function, in
order to promote the convergence of the cluster expansion 
used to calculate the energy expectation value \cite{clark79}.
In the low-density limit, the energy functional [binding energy 
per $\alpha$ cluster as a functional of the correlation factor 
$f(r)$] is given by 
\begin{equation}
E[f]= 2 \pi \rho_\alpha \int_0^\infty \left\{ \frac{\hbar^2}{m_\alpha}
    \, \left( \frac{\partial f(r) }{\partial r} \right)^2 +f^2(r)
  V_{\alpha}(r) \right\} r^2 dr \,,
\label{eev}
\end{equation}
where $m_\alpha$ is the $\alpha$-particle mass, while the condensate 
fraction is given by
\begin{equation}
\rho_0 = \exp \left\{-4 \pi \rho_\alpha \int_0^\infty [f(r) - 1]^2\,\,
  r^2 dr \right\}\,.
\end{equation}
The variational two-body correlation factor $f$ was taken
as one of the forms employed by Clark and coworkers \cite{JC80}, namely
\begin{equation}
f(r) = (1-e^{-ar})(1+be^{-ar}+ce^{-2ar})\,.
\end{equation}
At given density $\rho$, the expression for the energy expectation 
value is minimized with respect to the parameters $a$, $b$, and $c$, 
subject to the constraint (\ref{norm}).  It is important to note
that these approximations, based on truncated cluster expansions, 
are reliable only at densities low enough that the length scale 
associated with decay of $f^2(1)-1$ is sufficiently small compared 
to the average particle separation, which is inversely proportional 
to the cubic root of the density \cite{JC80,SMS06,clark79,Ristig}. 

\begin{figure}[htbp]
\begin{minipage}{8cm}
\begin{center}
\includegraphics[scale=0.32]{condfrac1.eps}
\caption{Reduction of condensate fraction in $\alpha$ matter with increasing
nucleon density. Exploratory calculations (full line) are compared with HNC
calculations of Johnson and Clark~\cite{JC80} (crosses). For comparison,
we show estimates of the condensate fraction in the $0_2^+$ (Hoyle)
state of $^{12}$C, according to Refs.~\cite{yamada1,suzuki} (stars).}
\label{fig:cond_fraction}
\end{center}
\end{minipage}
\hfill
\begin{minipage}{9.5cm}
\begin{center}
\includegraphics[scale=0.33]{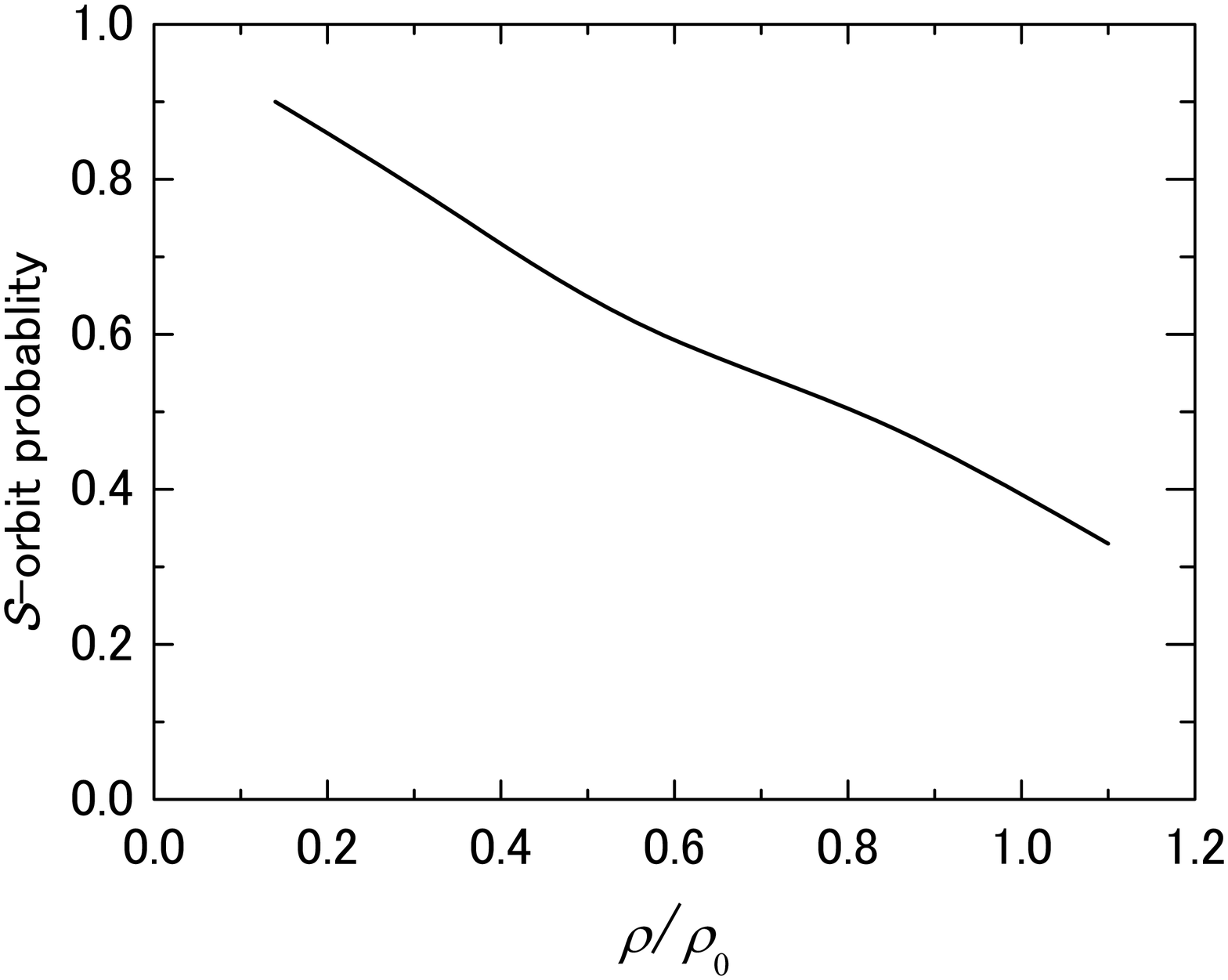}
\caption{Occupation of the $S$ orbital as a function of density using the $3\alpha$ OCM for $^{12}$C~\cite{yamada1}.}
\label{fig:cond_fraction_3alpha}
\end{center}
\end{minipage}
\end{figure}

To give an example, for the nucleon density $4 \rho_\alpha 
= 0.06$ fm$^{-3}$, a minimum of the energy expectation value
(\ref{eev}) was found at $a=0.616$ fm$^{-1}$, $b=1.221$, and $c=-5.306$, 
with a corresponding energy per $\alpha$ cluster of $-9.763$
MeV and a condensate fraction of 0.750.  The dependence of
the condensate fraction on the nucleon density $\rho = 4 \rho_\alpha$
as determined in this exploratory calculation is displayed in
Fig.~\ref{fig:cond_fraction}. 
\index{hypernetted chain methods}

The reduction of the condensate fraction of $\alpha$ matter to 
roughly 0.8 as given by our calculation
at nucleonic density 0.03 fm$^{-3}$ agrees well with results of
Suzuki \cite{suzuki} and Yamada \cite{yamada1} for $^{12}$C 
in the Hoyle $0^+_2$ state.  Using many-particle approaches to the 
ground-state wave function and to the THSR ($0_2^+$) state of 
$^{12}$C, the occupation of the inferred natural $\alpha$ orbitals 
is found to be quite different in the two cases. 
Roughly 1/3 shares (approaching equipartition) are found for the
$S$, $D$, and, $G$ orbits in the ground ($0_1^+$) state, 
with $\alpha$-cluster occupations of 1.07, 1.07, and 0.82, 
respectively.  On the other hand, in the Hoyle ($0^+_2$) state, 
one sees enhanced occupation (2.38) of the $S$ orbit and 
reduced occupation (0.29, 0.16, respectively) of the $D$ and 
$G$ orbits.  This corresponds to an enhancement of about 70\% 
compared with equipartition.

\begin{figure}[htbp]
\begin{center}
\includegraphics[scale=0.32]{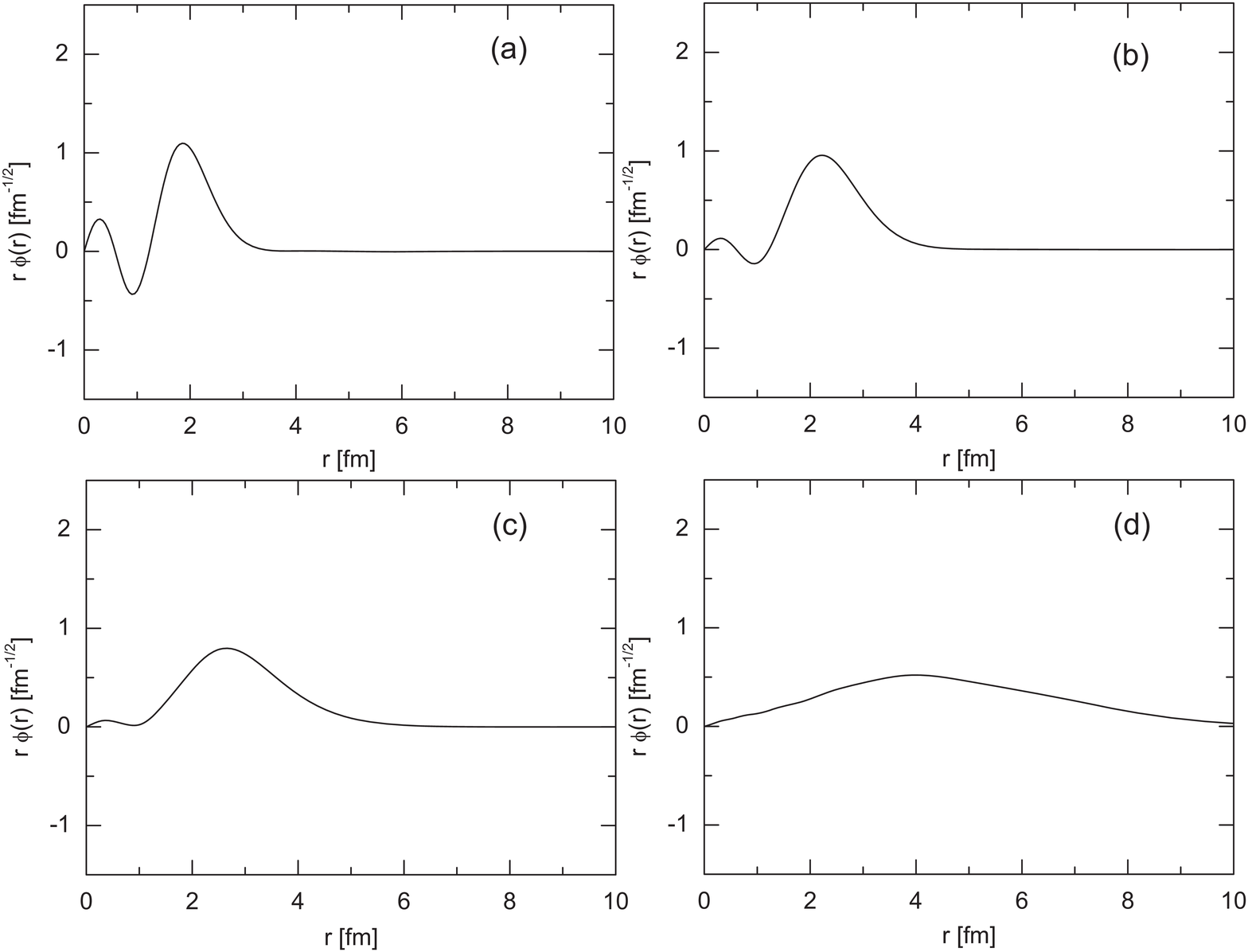}
\caption{Radial behaviors of the $S$ orbit in the $^{12}$C$(0^+)$ state with (a) $R = 2.42$ fm $(\rho/\rho_0\sim1.18)$, (b) $R = 2.70$ fm $(\rho/\rho_0\sim0.85)$, (c) $R = 3.11$ fm $(\rho/\rho_0\sim0.56)$, 
and (d) $R = 4.84$ fm $(\rho/\rho_0\sim0.15)$, where $R$ denotes the nuclear radius of the $^{12}$C$(0^+)$ state.}
\label{fig:13a-d}
\end{center}
\end{figure}

To get a more extended analysis, OCM calculations have been performed \cite{yamada1} for studying the density dependence of the $S$-orbit occupancy in the Hoyle state on the different densities $\rho/\rho_0 \sim (R{(0^+_1)}_{\rm exp}/R)^3$, in which the rms radius ($R$) of $^{12}$C is taken as a parameter and $R{(0^+_1)}_{\rm exp} $=2.56 fm. A Pauli-principle respected OCM basis $\Psi^{\rm OCM}_{0^+}(\nu)$ with a size parameter $\nu$ is used, in which the value of $\nu$ is chosen to reproduce a given rms radius $R$ of $^{12}$C, and the $\alpha$ density matrix $\rho(\vec{r},\vec{r}')$ with respect to $\Psi^{\rm OCM}_{0^+}(\nu)$ is diagonalized to obtain the $S$-orbit occupancy in the $0^+$ wave function. The results are shown in Fig.~\ref{fig:cond_fraction_3alpha}. The $S$-orbit occupancy is $70\sim 80$~\% around $\rho/\rho_0\sim (R{(0^+_1)}_{\rm exp}/R{(0^+_2)}_{\rm THSR})^3 = 0.21$, while it decreases with increasing $\rho/\rho_0$ and amounts to about $30\sim40$ \% in the saturation density region. Figure~\ref{fig:13a-d} shows the radial behaviours of the $S$-orbit with given densities. A smooth transition of the $S$-orbit is observed, with decreasing $\rho/\rho_0$, from a two-node $S$-wave nature $(\rho/\rho_0\sim 1.18)$ in Fig.~\ref{fig:13a-d}(a) to the zero-node $S$-wave one $(\rho/\rho_0\simeq0.15)$ in Fig.~\ref{fig:13a-d}(d) \cite{yamada1}. The feature of the decrease of the enhanced occupation of the $S$ orbit is in striking correspondence with the density dependence of the condensate fraction calculated for nuclear matter (see Fig.~\ref{fig:cond_fraction}).

 A more accurate and reliable variational description of $\alpha$
matter can be realized within the hypernetted-chain (HNC) approach 
to evaluation of correlated integrals; this approach \cite{JC80,clark79} 
largely overcomes the limitations of the cluster-expansion treatment,
including the need for an explicit normalization constraint.  Such
an improved approach is certainly required near the saturation 
density of nuclear matter, where it predicts only a small condensate 
fraction \cite{JC80}.  Of course, at high densities the simple 
Ali-Bodmer interaction \cite{AB66} ceases to be valid, and
it becomes crucial to include the effects of Pauli blocking.
Once again, this conclusion reinforces the point that we can expect
signatures of an $\alpha$ condensate only for dilute nuclei near 
the threshold of $n \alpha$ decay, but no signatures 
from configurations with saturated density.

\section{Conclusions, Discussion, Outlook}

We have investigated the role that pairing
and multiparticle correlations may play in nuclear matter existing 
in dense astrophysical objects and in finite nuclei.
A complete and quantitative description of nuclear matter
must allow for the presence of clusters of nucleons, 
bound or metastable, possibly forming a quantum condensate.
In particular, quartetting correlations, responsible 
for the emergence of $\alpha$-like clusters, are identified
as uniquely important in determining the behavior of nuclear 
matter in the limiting regime of low density and low temperature.
We have calculated the transition temperature for the onset 
of quantum condensates made up of $\alpha$-like and deuteron-like 
bosonic clusters, and considered in considerable detail the 
intriguing example of Bose-Einstein condensation of $\alpha$ 
particles. It turns out that contrary to pairing, quartet condensation 
primarily exists in the BEC phase at low density. In which way quartet 
condensation is lost by increasing the density is still an open question. 
However, it is clear that there can not exist a condensate of quartets with a 
long coherence length for arbitrarily small attraction as this is the case 
for pairing in the BCS phase.
It is inevitable that under increasing density or pressure,
the bound $\alpha$, $d$, or other nuclidic clusters 
present at low density experience significant modification due
to the background medium (and eventually merge with it).  
We have shown how self-energy corrections and Pauli blocking 
alter the properties of cluster states, 
and we have formulated a cluster mean-field approximation to 
provide an initial description of this process.  One result
of special interest is the suppression of the $\alpha$-like 
condensate, which is dominant at lower densities, as the density 
reaches and exceeds the Mott value, allowing the pairing 
transition to occur. Even at lower densities $\alpha$-particle condensation 
may be influenced by neutron excess, i.e. in the case of asymmetric nuclear 
matter. The study of $\alpha$-particle condensation as a function of 
asymmetry remains a task for the future.

 A truly remarkable manifestation of $\alpha$-particle condensation 
seems to be present in finite nuclei. Indeed, the so-called Hoyle 
state ($0_2^+$) in $^{12}$C at 7.654 MeV is very likely a 
dilute gas of three $\alpha$-particles, held together only by 
the Coulomb barrier.  This view is encouraged by the fact that we
can explain all the experimental data in terms of a conceptually 
simple wave function of the quartet-condensate type.  Within
the same model, we also systematically predict such states 
in heavier $n\,\alpha$ nuclei, and the search is on for their 
experimental identification. 
In a recent study with OCM (Orthogonality Condition Model) we predicted the 
6-th $0^+$ state of $^{16}$O to be a strong candidate for a loosely bound four 
$\alpha$-particle state~\cite{funaki_4aocm}. The results of that study are presented and 
discussed here. The condensate feature of the Hoyle state in $^{12}$C and 
Hoyle-like state in $^{16}$O is born out by the calculation of the bosonic 
occupation numbers in diagonalising the bosonic density matrix. It is 
shown that the occupation of the $0S$ state of the $\alpha$-particles is over 
70\% for the Hoyle state in $^{12}$C and over 60\% for the Hoyle-like state in 
$^{16}$O.

It is quite natural that such loosely bound $\alpha$-particle states should 
exist up to some maximum number of $\alpha$ particles.  We estimate that the phenomenon will terminate 
at about eight to ten $\alpha$'s as the confining Coulomb barrier 
fades away.  However, there is the possibility that larger 
condensates could be stabilized by addition of a few neutrons.  
Indeed, consider $^{9}$Be, which, contrary to $^{8}$Be, is bound by $\sim$ 
1.5 MeV, still showing a 
pronounced two $\alpha$-structure similar to the one of 
Fig.~\ref{profiles} (b). One could 
imagine ten $\alpha$'s or more, stabilised by two or four extra neutrons in a 
low density phase. However, even without being stabilised, if a compressed 
hot nuclear blob as e.g. produced in a central Heavy Ion collision expands and 
cools, it may turn on its way out, at a certain low density, into an 
expanding $\alpha$ condensed state 
where all $\alpha$'s are in relative $S$-waves. This would be an analogous 
situation to an expanding Bose condensate of atoms, once the trapping potential
has been switched off. Forthcoming analysis of dedicated experiments with high resolution 
multiparticle detectors like CHIMERA at LNS, Catania, will tell whether such 
scenarios can be realised \cite{Bord}. Other possibilities of loose 
$\alpha$-gas states may exist on top of particularly stable cores, like 
$^{16}$O or $^{40}$Ca. Indeed in adding $\alpha$'s to e.g. $^{40}$Ca, one 
will reach the $\alpha$-particle drip line. Compound states of heavy $N = Z$ 
nuclei of this kind may be produced in heavy ion reactions and an enhanced 
$\alpha$-decay rate may reveal the existence of an $\alpha$-particle 
condensate. Ideas of this type are presently promoted by von Oertzen 
\cite{koka,oertzen}, and also 
M. Brenner \cite{brenner}, and A. Ogloblin \cite{ogloblin}. However,  
coincidence measurements of 
multiple $\alpha$'s of decaying lighter nuclei like $^{16}$O may also be very 
useful \cite{freer} \cite{zarub} to detect at least one additional $\alpha$-condensate 
state 
beyond the only one that has been  identified so far, 
namely the $0_2^+$-state in $^{12}$C.

 In finite nuclei $\alpha$-particle condensation has to be understood in the 
same sense as we say nuclei are superfluid inspite of the presence of only a 
limited number of Cooper pairs. However, in compact stars, formed by 
Supernova Explosions, macroscopic condensates of $\alpha$-particles may be 
present. In the study of reference \cite{ST}, $\alpha$-particle phases are 
predicted for temperatures which can easily be below the critical 
temperatures obtained by our calculation displayed in Figs.~\ref{fig:trans_mu} and \ref{fig:trans_dens}. For the 
study of such macroscopic condensates, the use of the Gross-Pitaevskii equation 
for interacting ideal bosons can be useful \cite{yamada2} but may be hampered 
by our 
poor knowledge of the density dependence of the effective $\alpha$-$\alpha$ 
interaction. A genuine microscopic approach for $\alpha$-particle condensation 
in infinite matter is demanded, since our wave function, Eqs.~(\ref{eq:2}) 
and (\ref{eq:3}), only 
can handle a limited number of $\alpha$'s due to the necessary explicit 
antisymmetrisation. A preliminary route to this aim is outlined in this 
presentation.

Another issue which may be raised in the context of $\alpha$-particle 
condensation is the question, also discussed in condensed matter physics 
\cite{noz-jam}, whether $\alpha$'s condense as singles or as doubles, i.e. 
as $^{8}$Be? In microscopic studies of $^{12}$C one, indeed, can see that in 
the $0_2^+$-state two of the three $\alpha$'s are slightly more closer to 
one another than to the third one \cite{feld,enyo,neff}. 
Without Coulomb repulsion $^8$Be may be bound by 2-3 MeV, however, the Coulomb repulsion makes it very slightly 
unbound. The question is definitely very 
interesting and deserves future investigation. However, quantitatively, 
if at all, $\alpha$-$\alpha$ correlations constitute certainly only a slight modification over the 
present 
formulation of $\alpha$-condensation. One may even ask whether these 
$^{8}$Be-like 
correlations are not an artifact of the calculation because in the RGM 
calculation of \cite{kamimura} $^{8}$Be-like correlations are definitely 
allowed, 
however, as already mentioned, our pure $\alpha$-particle condensate wave 
function turned out to have almost 100 percent overlap with the one of the RGM 
calculation. 

 As already outlined earlier, nuclear systems exhibit especially strong 
cluster 
and few body effects. This stems primarily from the fact that there are four 
different fermions with in addition more or less equal attraction among one 
another. However, this situation is not necessarily unique. In the past there 
have been speculations that in semi-conductors bi-excitons may condense 
rather than excitons~\cite{nupecc}. A very promising field in this respect may be the 
possibility to trap in magneto-optical devices fermionic atoms in four 
different magnetic sub-states \cite{Salo}. If in addition the interaction 
between all four fermions could be triggered, eventually with the help of 
Feshbach resonances, so that they are all attractive, then investigation of 
quartetting could become a domain of research as rich as is presently the 
investigation of pairing. Clearly, the study of quartet condensation only 
is at its beginning and perspectives are rich and manyfold.

In conclusion, we see that the idea of $\alpha$-particle 
condensation in nuclei has already triggered many new ideas 
and calculations, in spite of the fact that, so far, a compelling 
case for such a state has only been made in $^{12}$C.  Even so,
the possible existence of a completely new nuclear phase in which
$\alpha$-particles play the role of quasi-elementary constituents is 
surely fascinating.  Hopefully, many more $\alpha$-particle states 
of nuclei will be detected in the near future, 
bringing deeper insights into the role of clustering and 
quantum condensates in systems of strongly interacting fermions.

\section*{Acknowledgements}
  We thank P. Nozi\`eres for his 
interest in quartet condensation, and we are greatful to J. W.~Clark, P.~Lecheminant and A.~Sedrakian for discussions and contributions.


\begin{thebibliography} {99}
\bibitem{foot}
 The value of the density of this state may be too low, 
since the extension of states around and above threshold depend very 
sensitively on their precise position.
\bibitem{4athsr}
The reason why we previously~\cite{nupecc} assigned the $0_4^+$ and $0_5^+$ states as being candidates for $\alpha$-condensation will be explained elsewhere.
\end{thebibliography}
\end{document}